\documentclass[11pt,a4paper,dvipsnames]{article}
\usepackage{cite}
\usepackage{amssymb}
\usepackage{amsmath}
\usepackage{amsthm}
\usepackage{xfrac}
\usepackage{latexsym}
\usepackage[dvips]{epsfig}
\usepackage{mathrsfs}
\usepackage{bm}
\usepackage{stmaryrd}
\usepackage{authblk}

\usepackage[dvipsnames]{xcolor}
\usepackage{tikz}
\usepackage{enumitem}
\usepackage{newtxtext}
\usepackage{array}

\theoremstyle{plain}

\newtheorem{definition}{Definition}

\allowdisplaybreaks

\def\bmd{{\bm d}}

\def\bmg{{\bm g}}
\def\bmh{{\bm h}}

\def\bml{{\bm l}}
\def\bmn{{\bm n}}
\def\bmm{{\bm m}}

\def\bmpartial{{\bm \partial}}

\usepackage{mathtools}% http://ctan.org/pkg/mathtools
\usepackage{xparse}% http://ctan.org/pkg/xparse
\makeatletter
\newcommand{\raisemath}[1]{\mathpalette{\raisem@th{#1}}}
\newcommand{\raisem@th}[3]{\raisebox{#1}{$#2#3$}}
\makeatother
\NewDocumentCommand{\newrbar}{O{0pt} O{0pt}}{
  \ensuremath{\mathrlap{\raisemath{#2}{\hspace*{#1}{\mathchar'26\mkern-9mu}}}r}}

\usepackage{amsfonts, amscd}

\usepackage{fullpage}
\newcounter{mnotecount}%[section]

\newcommand{\mnotex}[1]%{}
{\protect{\stepcounter{mnotecount}}$^{\mbox{\footnotesize $\bullet$\themnotecount}}$ 
\marginpar{%\color{red}%
\raggedright\scriptsize\em
$\!\!\!\!\!\!\,\bullet$\themnotecount: #1} }

\DeclareRobustCommand{\loongrightarrow}{%
  \DOTSB\relbar\joinrel\relbar\joinrel\rightarrow}
  
\newcounter{mnote}

\usepackage{color}
\usepackage{xcolor}
\usepackage[normalem]{ulem}
\usepackage{soul}
\usepackage{hyperref}

\begin{document}

\title{\textbf{ Asymptotic charges of a quadrupolar naked singularity}}

\author[1]{Edgar Gasper\'{i}n$^{*}$}
\author[2]{Mariem Magdy}

\affil[1]{Instituto de Ciencias Nucleares, Universidad Nacional Aut\'onoma 
de M\'exico\\ Cd.~Mx., 04510, M\'exico}
\affil[2]{Perimeter Institute for Theoretical Physics, 31 Caroline Street North, Waterloo, Ontario, N2L 2Y5, Canada.}

\date{}

\footnotetext[1]{* Corresponding author e-mail address: \texttt{e.gasperin@nucleares.unam.mx}}
\footnotetext[2]{\; E-mail address: \texttt{mmagdy@perimeterinstitute.ca}}

\maketitle
\begin{abstract}
The purpose of this article is to compute the asymptotic charges of a vacuum solution to the Einstein field equations describing a naked singularity with a non-vanishing quadrupole moment, known in the literature as the Zipoy–Voorhees spacetime ($q$-metric).
In addition to the well-known asymptotic quantities such as the Bondi–Sachs energy–momentum, the BMS charges and NP constants of this spacetime are computed.
Explicit calculations of the latter are relatively scarce in the literature. Moreover, it has been proven that the NP constants of asymptotically flat, stationary, vacuum, and algebraically special spacetimes vanish (for instance, those of the Kerr spacetime).
A by-product of the present analysis is to show that the algebraically special condition in the aforementioned result appears to be crucial, since the $q$-metric provides a counterexample to the conjecture that all asymptotically flat, stationary, vacuum, and asymptotically algebraically special spacetimes (a weaker version of the algebraically special condition) have vanishing NP constants.

\vskip .5em


\end{abstract}
\section{Introduction}

In the last decade, there has been a growing interest in the study of asymptotic structures and global quantities of isolated systems, such as the BMS charges, largely due to their relevance to black hole physics, gravitational waves, memory effects, and soft theorems \cite{Fav10,BlaDar92,Str14,HeLysMitStr15}. 
The BMS charges can be thought of as those quantities associated, via Noether's theorem, to the BMS group (Bondi, van der Burg, Metzner, and Sachs),
describing asymptotic symmetries at null infinity.
Being an infinite-dimensional group, there is an infinite number of these charges.  In general, the BMS charges are not conserved but rather sensitive to the flux of radiation through null infinity \cite{God20}. A particular example of one of these charges is the well-known Bondi mass, whose response to gravitational radiation is encoded in the Bondi mass loss formula. On the other hand, there exists a set of ten asymptotic quantities known as the Newman-Penrose (NP) constants which are, in contrast to the BMS charges,   conserved even in the presence of gravitational radiation.  Despite this importance conceptual difference, both the BMS charges and the NP constants are defined as certain integrals at cuts $\mathcal{C}$ of null infinity $\mathscr{I}$ and have been computed explicitly for some spacetimes, for instance for the Kerr spacetime in \cite{BarGlenTro11} and \cite{GonShaEtAl07}, respectively. In the case of  the Kerr spacetime, the BMS charges can be written in terms of the ADM mass $M$ and rotation parameter $a$. In contrast, the NP constants of the Kerr spacetime are zero. In fact, it has been proven in \cite{WuSha07} that the NP constants of asymptotically flat, stationary and algebraically special spacetimes are zero. Hence, it is not surprising that the NP constants of the Kerr spacetime vanish as it is, besides stationary and asymptotically flat, a Petrov-type D spacetime.

\medskip

Another explicit example of the calculation of NP constants was provided in \cite{LazVal00} where the NP constants of Type D boost-rotation spacetimes (including, in particular, the C-metric) were found to be generally non-vanishing. This is not in conflict with the theorem in \cite{WuSha07}, as these spacetimes, although algebraically special, are not stationary. In \cite{WuSha07}, it was observed that the algebraically special condition was only used asymptotically and it was conjectured that the theorem could hold for a larger class of spacetimes; those which are only asymptotically algebraically special. In this article, we show, by means of an explicit example, that the algebraically special condition appears to play a critical role, as the NP constants of an asymptotically flat, stationary but algebraically general vacuum solution ---known as the Zipoy–Voorhees spacetime--- are, in fact, non-zero, despite being “asymptotically algebraically special”. The Zipoy-Voorhees (also known as the $q$-metric) spacetime is an exact static axisymmetric solution to the vacuum Einstein field equations which can be thought of as a simple generalisation of the Schwarzschild spacetime, allowing for a non-vanishing quadrupole moment and containing a naked singularity \cite{PapStewWitt81}. Although the static, vacuum and axisymmetric solutions to the Einstein field equations can be written in terms of infinite series for two metric potentials by solving linear partial differential equations (PDEs), in most cases, the metric cannot be written in explicit and simple closed form. Two particular examples where this can be done are the Schwarzschild metric and the so-called $q$-metric. The $q$-metric  contains two parameters $m$ and $q$ related to the mass and quadrupole moment. If one sets $q=0$ and $q=-1$, one recovers the Schwarzschild and  Minkowski spacetimes, respectively ---see \cite{Que11} and \cite{HerFilSan99} for other limits of this spacetime. Since the $q$-metric can be written in a compact closed form, it has been used to study integrability, astrophysical effects and geodesic motion in the presence of naked singularities --- see e.g. \cite{Que11,BosGasGutQueTor16,DanUteQueUra25,SerKuaQueEtAl24,BosKonEtAl21,HerPaiFilSan98,Lukes05,Mal04,ChAniPatMalJos11,HerBar04}.
The purpose of this paper is to study the $q$-metric from the perspective of its asymptotic properties by deriving formal asymptotic  expansions for the metric and the associated null tetrad in the Newman-Penrose (NP) gauge. This, in turn, allows for the computation of asymptotic charges such as the Bondi mass, BMS charges, and NP constants.

\medskip

This article is structured as follows: Section \ref{Sec:global} 
 introduces the
$q$-metric and
its main properties and, crucially for this article, its Petrov-type is analysed. In Section \ref{Sec:Asymptotics-NP-Gauge}, the asymptotic form of the $q$-metric is obtained in the NP-gauge by means of asymptotic expansions  close to the conformal boundary  and explicit expressions for the spin-coefficients and Weyl scalars in the NP-gauge are derived. In Section \ref{Section:Asymptotic-conserved-quantities}, explicit expressions for the Bondi-Sachs energy-momentum along with the NP constants and the BMS charges are obtained.  Section \ref{Sec:Conclusions} contains the conclusions of this work.

\section{The Zipoy-Voorhees spacetime and its Petrov-type}
This section gives a concise introduction to the Zipoy-Voorhees spacetime and its main features.\label{Sec:global}
\subsection{The \texorpdfstring{$q$}{q}-metric}
The Zipoy-Voorhees spacetime is a member of a general class of solutions to Einstein's field equations describing static and axisymmetric spacetimes ---the so-called Weyl metric. The Weyl metric \cite{Weyl17} can be written in terms of Weyl coordinates $(t,\varrho,z,\phi)$ as 
\begin{equation}\label{eq:Weyl_metric_form}
    \bmg_{W} =  e^{-2 \psi} \bmd{t} \otimes \bmd{t} - e^{2(\gamma-\psi)} (\bmd{\varrho} \otimes \bmd{\varrho} + \bmd{z} \otimes \bmd{z}) - e^{-2 \psi} \varrho^2 \bmd{\phi} \otimes \bmd{\phi},
\end{equation}
where $\psi := \psi(\varrho,z)$ and $\gamma :=\gamma(\varrho,z)$ are the two metric potentials.
The vacuum Einstein field equations $R_{ab}=0$, reduce to a linear equation for $\psi$. Once this equation is solved, $\gamma$ can be obtained by quadratures \cite{GriPod09}. This allows for obtaining explicit solutions and even generating new solutions from already known ones. In particular, the Zipoy-Voorhees transformation was used in \cite{Que11} to generate a vacuum solution which is a generalisation of the Schwarzschild metric that was called the $q$-metric in subsequent works ---see for instance \cite{BosGasGutQueTor16, DanUteQueUra25, BosKonEtAl21}. In Weyl coordinates, the $q$-metric corresponds to the following potentials $\psi=\psi_{q}$ and $\gamma=\gamma_{q}$, given in terms of two constant parameters $m$ and $q$ by
\begin{equation}
    \psi_{q} = \frac{1+q}{2} \log{\frac{L-m}{L+m}} , \qquad \gamma_{q} = \frac{(1+q)^2}{2} \log{\frac{L^2-m^2}{l_{-}l_{+}}},
    \label{Potentials}
\end{equation}
where
\begin{equation*}
    L = \frac{1}{2} (l_{+}+l_{-}), \qquad l_{\pm} = \sqrt{\varrho^2+(z \pm m)^2}.
\end{equation*}
Setting $q=0$ retrieves the Schwarzschild solution.
To see this more clearly and to write the $q$-metric in a more recognisable form, one
introduces coordinates
\begin{equation}
    \varrho = \sqrt{r^2 - 2mr + q^2} \sin{\theta}, \qquad z = (r-m) \cos{\theta},
\end{equation}
so that, in the $(t,r,\theta,\phi)$ coordinates the $q$-metric reads
\begin{eqnarray}
    && \bmg =  F^{1+q} \bmd{t} \otimes \bmd{t} - F^{-1-q} G^{-q (2+q)} \bmd{r} \otimes \bmd{r} \nonumber \\
    && \phantom{\bmg_{q}} - r^2 F^{-q} G^{-q (2+q)} \bmd{\theta} \otimes \bmd{\theta} - r^2 F^{-q} \sin^2{\theta} \bmd{\phi} \otimes \bmd{\phi}, \qquad \label{$q$-metric}
\end{eqnarray}
where
\begin{equation}
    F:=1-\frac{2m}{r} \qquad G:=1 + \frac{m^2\sin^2 \theta }{r^2-2m r}. \label{Definition-F-G}
\end{equation}
This metric is an axially-symmetric exact solution to Einstein's vacuum equations $R_{ab}=0$. In its form above, it is clear that the Schwarzschild metric is a special case for which $q=0$. In fact, the discussion in \cite{BosGasGutQueTor16} shows that $q$ determines the quadrupole moment since the lowest Geroch-Multipole-moments \cite{Geroch70} $M_{n}$ can be written as 
\begin{equation*}
    M_{0} = (1+q)m, \quad M_{2} = -\frac{m^3}{3} q(1+q)(2+q), \ldots
\end{equation*}
---see \cite{Que11, FruQueSan18} for a detailed calculation.
The $q$-metric spacetime exhibits curvature singularities at $r=0$, $r=2m$, and along the surface defined by
\begin{equation*}
r^2 - 2mr + m^2 \sin^2 \theta = 0, \qquad q \in \left(-1, -1 + \sqrt{\tfrac{3}{2}} \right) \setminus \{0 \}.
\end{equation*}
This can be seen by analysing the Kretschmann scalar $K$, which for the $q$-metric is given by \cite{Que11}
\begin{equation} K= \frac{16m^2 (1+q)^2 (r^2-2mr+m^2\sin^{2}\theta)^{2(2q+q^2)-1}}{r^{4(2+q+q^2)} (1-\tfrac{2m}{r})^{2(q^2+q+1)}} L(r,\theta), \label{Kretschmann-scalar} 
\end{equation}
where $L(r,\theta)$ is
\begin{eqnarray*} && L(r,\theta) = 3 (r-2m - qm)^2 (r^2 - 2mr +m^2 \sin^2 \theta), \\ && \phantom{L(r,\theta)} + q (2+q) (q (2+q) + 3 (r-m)(r-2m-qm)) \sin^2 \theta. 
\end{eqnarray*}
The divergence of $K$ as $r \to 0$, $r \to 2m$, and along the zeros of $r^2 - 2mr + m^2 \sin^2 \theta$ follows directly from the structure of \eqref{Kretschmann-scalar}. In particular, for all real $q$ one has $2+q+q^2>0$ and $q^2+q+1>0$, implying that $K \to \infty$ as $r \to 0$ and $r \to 2m$.

Moreover, the exponent $2(2q+q^2)-1<0$ for $q \in \left(-1,-1+\sqrt{\tfrac{3}{2}}\right)$. For $q=0$, one finds
\begin{equation*}
L(r,\theta) = 3 (r-2m)^2 (r^2-2mr - m^2 \sin^2 \theta),
\end{equation*}
so the potential divergence is cancelled. However, for $q \in \left(-1, -1 + \sqrt{\tfrac{3}{2}} \right) \setminus \{0 \}$,
\begin{equation*}
(r^2-2mr+m^2\sin^{2}\theta)^{2(2q+q^2)-1} L(r,\theta) \to \infty \qquad \text{as } r \to m \pm m \cos\theta,
\end{equation*}
showing that this surface corresponds to a genuine curvature singularity.
Notice that, although the Killing vector $\bmpartial_t$ becomes null at $r=2m$, the divergence of $K$ there indicates that this is a singular hypersurface rather than an  horizon. Hence, the $q$-metric describes a spacetime containing a naked singularity.

\medskip

Motivated by this fact, different aspects of this geometry have been analysed to find astrophysical features that could signal the presence of a naked singularity: for instance, through the structure of its accretion disk \cite{BosGasGutQueTor16, Far22}. 
This spacetime is interesting not only for its astrophysical applications but also for its asymptotic properties, which we investigate in this article.
                           
\subsection{Petrov-type of the \texorpdfstring{$q$}{q}-metric}
Many of the well-known explicit (exact) vacuum solutions of the Einstein field equations are \emph{algebraically special}. An interesting feature of the $q$-metric, as we show in this section, is that despite its simplicity, it is \emph{algebraically general}. To see this, observe that a null tetrad for the $q$-metric
\eqref{$q$-metric} is given by
\begin{subequations}
    \begin{eqnarray}
        && \sqrt{2}\; \mathring{\bm l} =  F^{-(1+q)/2} \;\bm\partial_{t} + F^{(1+q)/2}G^{q(2+q)/2}\bm\partial_{r}, \\
        && \sqrt{2}\; \mathring{ \bm n} = F^{-(1+q)/2}\;\bm\partial_{t} - F^{(1+q)/2}G^{q(2+q)/2}\;\bm\partial_{r},\\
        && \sqrt{2}\;  r \; \mathring{ \bm m} = F^{q/2}G^{q(2+q)/2}\;\bm\partial_{\theta} + \mathbf{i} \;F^{q/2}\csc\theta \; \bm\partial_{\phi}.
    \end{eqnarray}
    \label{Original-null-tetrad}
\end{subequations}
The normalisation and signature are chosen so that
$\bmg (\mathring{\bm l}, \mathring{\bm n})= -\bmg (\mathring{\bm m}, \bar{\mathring{\bm m}})= 1$ while all the other contractions vanish. Since later we will consider a Lorentz transformation of this basic tetrad, we have added a decoration $\mathring{\cdot}$ to denote this simple choice of tetrad and all the quantities derived from it. A direct calculation  gives the spin-coefficients and the Weyl scalars of the NP formalism. To ensure the flow of the discussion these quantities are given in Appendix \ref{Apendix:Connection-Weyl-NP}.
From the expressions in equation \eqref{Original-tetrad-Weyl-components} one could naively try to read off some physically relevant quantities by noticing that
\begin{subequations}
    \begin{eqnarray}
    && \mathring{\Psi} _{0}{} = \frac{m^3 q (1 + q) (2 + q) \sin^2\theta }{2 r^5} + \frac{m^3 q (1+q) (2+q) (2 q-3) \sin^2{\theta}}{2 r^6} +\mathcal{O}(r^{-7}),\\
    && \mathring{\Psi} _{1}{} = - \frac{m^3 q (1 + q) (2 + q) \sin(2\theta) }{4 r^5} + \frac{m^4 q (1+q) (2+q) ( 2 q - 3) \sin{(2\theta)}}{4 r^6} + \mathcal{O}(r^{-7}),\\
    && \mathring{\Psi}_{2} = - \frac{m (1 + q)}{r^3} + \frac{3 m^2 q (1+q)}{r^4} + \frac{m^3 q (1+q) (10-19q+3(2+q) \cos(2 \theta))}{4 r^5} \nonumber\\
    && \phantom{\mathring{\Psi} _{2}}+ \frac{m^4 q (1+q) (-38+(93-64q)q + 3(2+q) (-7+8q) \cos(2 \theta))}{12 r^6} +\mathcal{O}(r^{-7}),\\
    && \mathring{\Psi} _{3}{} = \frac{m^3 q (1 + q) (2 + q) \sin(2\theta) }{4 r^5} - \frac{m^4 q (1+q) (2+q) (2q-3) \sin(2 \theta)}{4 r^6} + \mathcal{O}(r^{-7}),\\
    && \mathring{\Psi} _{4}{} = \frac{m^3 q (1 + q) (2 + q) \sin^2\theta }{2 r^5} - \frac{m^4 q (1+q) (2+q) (2q-3) \sin^2{\theta}}{2 r^6} +\mathcal{O}(r^{-7}).
    \end{eqnarray}
    \label{original-Weyl-components}
\end{subequations}
Observe that the Coloumb's component $\mathring{\Psi}_2$ correctly encodes the ADM mass which, according to the detailed discussion in the Appendix of \cite{BosGasGutQueTor16}, is given by $M_{ADM} = m (1+q)$. However, in general, the simple tetrad $(\mathring{\bm l}, \mathring{\bm n} , \mathring{\bm m}, \bar{\mathring{\bm m}})$ and the coordinates $(t,r,\theta, \phi)$ do not correspond to those used in the \emph{theory of asymptotics of Newman and Penrose} ---or that of Bondi and Sachs--- to define physically meaningful quantities \cite{NewPen62,BMS62}. Fortunately, to determine whether the spacetime is algebraically special, one can use the following curvature invariants ---hence, independent of the choice of coordinates and tetrad:
\begin{equation*}
    I = 6 \mathring{\Psi} _{2}{}^2  -8 \mathring{\Psi} _{1}{} \mathring{\Psi} _{3}{} + 2 \mathring{\Psi} _{0}{} \mathring{\Psi} _{4}{} \qquad
J = 6 (- \mathring{\Psi} _{2}{}^3 + 2 \mathring{\Psi} _{1}{} \mathring{\Psi} _{2}{} \mathring{\Psi} _{3}{}  - \mathring{\Psi} _{0}{} \mathring{\Psi} _{3}{}^2  - \mathring{\Psi} _{1}{}^2 \mathring{\Psi} _{4}{} + \mathring{\Psi} _{0}{} \mathring{\Psi} _{2}{} \mathring{\Psi} _{4}{}).
\end{equation*}
A spacetime is said to be algebraically special if $P:=I^3 -6 J^2 =0$ ---see \cite{PenRin86}. For the case of the $q$-metric, a direct   calculation gives
\begin{align*}
    P=\frac{G^{6q(2+q)}\;\Gamma\; \Delta}{\Sigma},
\end{align*}
where 
\begin{align*}
    &\Gamma =  3 r (2 m^3 (2 + q) (-15 -2 q (6 + q) + (3 + 2 q (3 + q)) \cos(2 \theta )) + m^2 (75 + 2 q (28 + 5 q) \\ & \qquad
    - (3 + 4 q (2 + q)) \cos(2 \theta )) r  -12 m (3 + q) r^2 + 6 r^3) + 2 m^4 (2 + q)^2 (3 + 2 q)^2 \sin^2\theta, \\
    &\Delta = m^{10} q^2 (1 + q)^6 (2 + q)^2F^{6 q} (m^2 (2 + q) (3 + q) -3 m (3 + q) r + 3 r^2)^2 \sin^4\theta, \\
    & \Sigma = r^{18} (r- 2 m )^6 (r (r-2m) + m^2 \sin^2\theta )^3.
\end{align*}
Hence, it is clear that the spacetime is algebraically general ---or type I in the Petrov classification \cite{Pet54}. Recall that Petrov type I corresponds to the algebraically general case of the Weyl tensor, in which all four principal null directions are distinct and no special algebraic symmetries are present. However, these principal null direction do not necessarily span the four-dimensional space ---for further discussion on the Petrov classification and its implications see \cite{StepameMacHoeHer03}. Also, observe that  $P=0$ for $q=-1$ and $q=0$ is consistent with the fact that these cases correspond to the Minkowski and the Schwarzschild spacetimes,  which are, respectively, of Petrov-type O and D.

\medskip
The algebraically special condition is satisfied by most of the well-known electro-vacuum spacetimes (PP-waves, Schwarzschild, Reissner-Nordstroem, Kerr, Kerr-Newman etc.) and important consequences follow from this condition in terms of integrability of their geodesic motion. For instance, Petrov-type D spacetimes (such as the Kerr spacetime) admit hidden symmetries (Killing spinors) which are ultimately responsible for the existence of the Carter constant \cite{WalPen70}.  Therefore, a spacetime which is alebraically general will not admit a Killing spinor and hence there will not admit a conserved quantity analogous to the Carter constant.
Consequently, the fact that the Zipoy-Voorhees spacetime is algebraically general is consistent with the observation made in \cite{MacPrzSta13} that geodesic motion (outside of the equatorial plane) is not  Liouville-integrable ---see also \cite{Des23}. Nonetheless, notice that, asymptotically, for the $q$-metric one has:
\[
P=\frac{162 m^{10} q^2 (1 + q)^6 (2 + q)^2 \sin^4\theta }{r^{22}} + \mathcal{O}(r^{-23}).
\]
In \cite{WuSha07, ZhaXiaGao09}, it was shown that any asymptotically flat and stationary spacetime satisfies $P = \mathcal{O}(\mathfrak{r}^{-21})$ near null infinity, and hence a spacetime was defined to be \emph{asymptotically algebraically special} if $P$ decays one order faster:  $P = \mathcal{O}(\mathfrak{r}^{-22})$. In these definitions, the coordinate $\mathfrak{r}$ denotes an affine parameter along outgoing null geodesics ---see \cite{NewUnt62} for a discussion on the NP-gauge.  In the following section, it will be shown that, for the $q$-metric, to leading order, one has $r = \mathfrak{r} + \mathcal{O}(\mathfrak{r}^{-1})$ ---see equation \eqref{Explicit-form-spherical-r}. Therefore, $P= \mathcal{O}(\mathfrak{r}^{-22})$ and then the Zipoy-Voorhees spacetime is an example of an asymptotically algebraically special solution. However, as we show in this article, the NP constants of the Zipoy–Voorhees spacetime are non-vanishing. This indicates that the assumption of algebraic speciality in the theorem proved in \cite{WuSha07} likely cannot be relaxed. Although \cite{ZhaXiaGao09} also claims to establish the result under the relaxed assumption, our findings suggest that its validity may depend on additional, implicit assumptions—possibly related to spacetime singularities, its global structure, or the relaxation of algebraic speciality itself.

\medskip

A few remarks on asymptotic flatness and peeling for vacuum solutions in the Weyl class are now in order. Consider a vacuum solution \((\mathcal{M}, \bmg_W)\) of the Einstein field equations in the Weyl class, with \(\bmg_W\) given by \eqref{eq:Weyl_metric_form}. Let \(\mathcal{S}\) denote the \(t=0\) hypersurface, and let \(\bmh\) be the metric induced on \(\mathcal{S}\) by \(\bmg_W\). As shown in \cite{Geroch70}, the initial data set \((\mathcal{S}, \bmh)\) is an asymptotically Euclidean and regular manifold; see Appendix~\ref{Ap:AsymptEuclidean} for definitions and for an explicit computation in the Zipoy--Voorhees spacetime.
Furthermore, \cite{Dai01} establishes that every stationary, asymptotically flat vacuum spacetime admits an analytic conformal extension through null infinity ($\mathscr{I}$). Since the \(q\)-metric is a static vacuum solution within the Weyl class, these results imply that although the Zipoy--Voorhees spacetime has a naked singularity and hence, \(\mathscr{I}\) is not complete, it nevertheless admits a portion of \(\mathscr{I}\), close to spatial infinity (\(i^0\)), which is smooth.
The smoothness of the conformal boundary plays a crucial role: it guarantees the validity of the peeling behaviour, ensuring that the  surface integrals defining the  Newman--Penrose constants and the BMS charges are well defined. See \cite{MagPrabVal24, MagVal22, GasMagMen25, GasPin23, Val98-2} for further discussion on the importance of peeling (and the consequences of its failure) for the existence of asymptotic charges.

\medskip
                        
%-----------------------------------------------------------------------                                                                                                                                           
\section{Asymptotics in the Newman-Penrose gauge}
\label{Sec:Asymptotics-NP-Gauge}
%-----------------------------------------------------------------------    

In this section, asymptotic expansions for the NP tetrad and NP coordinates for the Zipoy-Voorhees spacetime are obtained.
The NP-gauge is concisely introduced in subsection \ref{NPgaugeDefs}. The relation between the original and the NP coordinates is given in subsection \ref{NP-coordinates}.
The Lorentz transformation relating the original and the NP frame is obtained in subsection \ref{LorentzTransf}. The final expressions for the spin-coefficients and Weyl scalars in the NP-gauge are given in subsection \ref{Subsec:NP-quantities-NP-coordinates}.

\subsection{The NP-gauge}\label{NPgaugeDefs}
In the seminal work of Newman and Penrose (NP) \cite{NewPen62} ---see also \cite{NewUnt62, NewPen68}, special coordinates
and a null frame well adapted for the study of asymptotics
were introduced. 
To construct the NP-gauge on an asymptotically flat spacetime, one begins by considering a solution $u$ to the \emph{eikonal} equation
\begin{align}
\bmg^{\sharp}(\bm{d}u, \bm{d}u) = 0,
\label{Eikonal-equation}
\end{align}
which ensures that $u$  defines a family of null hypersurfaces. These hypersurfaces, denoted by $ \mathcal{N}_u$, form a foliation of the spacetime corresponding to level sets of constant  $u$.
The null vector $\bm l^{\sharp} = \bm d u$  is
chosen so that it is tangent to a family of null geodesics, namely
$\nabla_{\bm  l} \bm l=0$. On each of these null geodesics, the affine
parameter $\mathfrak{r}$ is used as a  coordinate. Finally,
at some fiduciary cut $\mathcal{C} \subset \mathscr{I} \approx  \mathbb{S}^2$, one considers some coordinates $x^A$ with $A=2,3$ ---usually taken to be the
standard $(\theta, \varphi)$ angular coordinates on $\mathbb{S}^2$--- which are required to remain
constant along the generators of $\mathscr{I}$ and on outgoing null
geodesics. In the NP-coordinates $(u,\mathfrak{r}, x^A)$ the adapted frame reads
\begin{subequations}
    \begin{eqnarray}
    && \bm l = \bm\partial_{\mathfrak{r}}, \\
    && \bmn =  \bm\partial_u + U\bm\partial_\mathfrak{r} + X^A\bm\partial_{A}, \\
    && \bm m = \omega \bm\partial_{\mathfrak{r}}+\xi^A\bm\partial_{A},
    \end{eqnarray}
\end{subequations}
for some scalar functions $\{U, X^A,\omega, \xi^A\}$ effectively encoding  the metric components.
The remaining freedom in the frame is fixed by requiring $\bm m$
and $\bm n$ to be parallelly propagated along $\bm l$ which translates
into the following conditions for the spin-coefficients --- see Appendix B in \cite{Stewart91} for their definition:
\begin{align}
  \kappa = \pi = \epsilon =0, \quad \rho = \bar{\rho}, \quad \tau = \bar{\alpha} + \beta.
  \label{NP-gauge-conditions}
\end{align}
Note that for the $q$-metric, the condition $\mathring{\rho} = \mathring{\bar{\rho}}$ is automatically satisfied since all spin-coefficients associated to the tetrad $(\mathring{\bml},\mathring{\bmn},\mathring{\bmm},\mathring{\bar{\bmm}})$, are real ---see equations \eqref{eq:RingSpinCoef}. The other conditions can be enforced through a suitable Lorentz transformation.

\subsection{NP-coordinates}
\label{NP-coordinates}
The first step in the construction is to find a solution to the eikonal equation \eqref{Eikonal-equation} that will  define the \emph{retarded time} coordinate $u$. Motivated by the fact that the components of the $q$-metric \eqref{$q$-metric} are independent of the $\phi$ coordinate, for simplicity, we take $u = u(t,r,\theta)$. Then, a straightforward calculation shows that the eikonal equation for the $q$-metric reads
\begin{equation*}
    \left( 1- \frac{2m}{r} \right)^{-1-q} (\partial_{t}u)^2 - \left( 1- \frac{2m}{r} \right)^{q} \left( 1+ \frac{m^2 \sin^2{\theta}}{r^2-2mr}\right)^{q (2+q)} \left( \left( 1- \frac{2m}{r} \right) (\partial_{r} u)^2 + \frac{(\partial_{\theta} u)^2}{r^2} \right)=0.
\end{equation*}
A formal solution to this equation can be obtained in the asymptotic region  %$r \to \infty$ 
by considering the Ansatz 
\begin{equation}
    u = t -r -2m (1+q) \log{\frac{r}{2m(1+q)}} + \sum_{k=1}^{\infty} \frac{f^{(k)}(\theta)}{r^k},
    \label{u-Ansatz}
\end{equation}
where the upper index in $f^{(k)}$ does not denote derivatives, it simply labels a collection of functions $\{f^{(1)}, f^{(2)}, \ldots\}$. This notation, instead of the simpler $f_k$, is used  for consistency with the upcoming asymptotic expansions for quantities with sub-indices such as the Weyl scalars. 
Substituting this Ansatz, $f^{(k)}(\theta)$ can be determined up to any desired order.
For instance, to order $k\leq 2$ one gets
\begin{flalign*}
         f^{(1)} (\theta) &= \frac{1}{4} m^2 (2+q) (8+7q+q \cos(2 \theta)), \\ 
       f^{(2)}(\theta) &=  \frac{1}{12} m^3 (2+q) (24+26q+5q^2+3q(2+q)\cos(2 \theta)).
\end{flalign*}
Namely, to this order in the expansion, one can write:
\begin{equation}
    u = t-r-2m (1+q) \log{\frac{r}{2m(1+q)}} + \frac{m^2(2+q) (8+7q+q \cos(2 \theta))}{4r} + \mathcal{O}(r^{-2}). 
    \label{asymptotic-u-solution}
\end{equation}
Using formal expansions is sufficient, as only relatively low orders are required to accurately compute the asymptotic charges.  
Therefore, questions regarding the convergence of the formal solution lie beyond the scope of this paper.  
The asymptotic expansions for the components of \(\bmg\) and its inverse \(\bmg^\sharp\) in the coordinates \((u, r, \theta, \phi)\) can be readily computed and are provided in Appendix~\ref{Appendix:CoordinatesExpansions}.  
With these expressions at hand, we proceed to determine the NP coordinate \(\mathfrak{r}\).  
To this end, recall that the NP-gauge requires \(\bml^\flat = \bmd u\) and \(\bml = \bm\partial_{\mathfrak{r}}\).  
Exploiting the asymptotic expansions of the metric, and noting that \(\bm\partial_{\mathfrak{r}} = \bmg^\sharp(\bmd u, \cdot)\), one obtains
\begin{equation}    \bmpartial_{\mathfrak{r}} = \bmpartial_{r} -\frac{m^2 q (2+q) \sin^2 \theta}{2r^2} \bmpartial_{r} - \frac{m^2 q (2+q)(\cos\theta \sin\theta \bmpartial_{\theta} + m \sin^{2}\theta \bmpartial_{r})}{r^3} + \mathcal{O}(r^{-4}). 
    \label{Affine-parameter-equation}
\end{equation}
Since we are only interested in the asymptotic region, proceeding as before, we take the Ansatz
\begin{equation}
    \mathfrak{r} = r +\sum_{k=1}^{\infty} \frac{h^{(k)}(\theta)}{r^k}.
    \label{ansatz-affine-parameter}
\end{equation}
Solving to order $k \leq 2$ for  $h^{(k)}(\theta)$ gives
\begin{equation*}
    h^{(1)}(\theta) = \frac{1}{2} m^2 q (2+q) \sin^{2}\theta, \qquad h^{(2)}(\theta) =  \frac{1}{2} m^3 q (2+q) \sin^{2}\theta,
\end{equation*}
which guarantees that equation \eqref{Affine-parameter-equation} is satisfied up to the expanded order. In general, the expressions for $h^{(k)}(\theta)$, can be determined up to a desired order provided the metric components are sufficiently expanded in the $(u,r,\theta,\phi)$ coordinates ---see Appendix \ref{Appendix:CoordinatesExpansions}.
The inverse transformation $r = r (\mathfrak{r},\theta)$ can obtained by taking a similar Ansatz and substituting into equation \eqref{ansatz-affine-parameter}. To the given order, one finds,
\begin{equation}
    r = \mathfrak{r} - \frac{m^2 q (2+q) \sin^{2}\theta}{2 \mathfrak{r}} - \frac{m^3 q (2+q) \sin^{2}\theta}{2 \mathfrak{r}^2} + \mathcal{O}(\mathfrak{r}^{-3}).
    \label{Explicit-form-spherical-r}
\end{equation}
The fall-off of the metric components in NP coordinates $(u,\mathfrak{r},\theta,\phi)$ can be found in equation \eqref{eq:MetricInNPcoords} of Appendix \ref{Appendix:CoordinatesExpansions}.

\subsection{Lorentz transformations}\label{LorentzTransf}
\subsubsection*{Decomposed transformation}
As previously discussed, the NP-gauge involves not only a specific choice of coordinates but also fixing tetrad such that the conditions \eqref{NP-gauge-conditions} are satisfied. Naturally, the original tetrad $( \mathring{\bml}, \mathring{\bmn},\mathring{\bmm}, \mathring{\bar{\bmm}} )$ and the targeted tetrad $( \bml, \bmn, \bmm, \bar{\bmm})$ are related by a Lorentz transformation.
In this case, the desired Lorentz transformation can be obtained as the composition of three successive, simpler transformations:
\begin{equation*}
    (\mathring{\bml}, \mathring{\bmn}, \mathring{\bmm}) 
    \overset{\Lambda_1}{\loongrightarrow} 
    (\bml'', \bmn'', \bmm'') 
    \overset{\Lambda_2}{\loongrightarrow} 
    (\bml', \bmn', \bmm') 
    \overset{\Lambda_3}{\loongrightarrow} 
    (\bml, \bmn, \bmm).
\end{equation*}
Here, \(\Lambda_1\) denotes a null rotation about \(\mathring{\bmn}\), employed to impose the condition \(\kappa'' = 0\);  
\(\Lambda_2\) represents a null rotation about \(\mathring{\bml}\), used to set \(\pi' = 0\);  
and \(\Lambda_3\) corresponds to a spin-boost transformation, applied to enforce \(\epsilon = 0\) and \(\tau = \bar{\alpha} + \beta\).
Each of these transformations is chosen so that each step preserves the conditions imposed by the previous ones. For example, once $\kappa'' = 0$ is achieved through $\Lambda_1$, the application of $\Lambda_2$ ensures that $\kappa' = 0$, and after applying $\Lambda_3$, one still has $\kappa = 0$.
\medskip
Explicitly, these transformations are:
\[
\begin{array}{@{}l@{\quad}l@{}}
\Lambda_1:\ 
\left\{
\begin{array}{l}
\bml'' := \mathring{\bml} + a_1 \mathring{\bmm} + a_1 \bar{\mathring{\bmm}} + a_1^2 \mathring{\bmn}, \\
\bmn'' := \mathring{\bmn}, \\
\bmm'' := \mathring{\bmm} + a_1 \mathring{\bmn}.
\end{array}
\right.
& 
\Lambda_2:\ 
\left\{
\begin{array}{l}
\bml' := \bml'', \\
\bmn' := \bmn'' + a_2 \bmm'' + a_2 \bar{\bmm}'' + a_2^2 \bml'', \\
\bmm' := \bmm'' + a_2 \bml''.
\end{array}
\right.
\\[2em]
\Lambda_3:\ 
\left\{
\begin{array}{l}
\bml := a_3 \bml', \\
\bmn := \bmn'/a_3, \\
\bmm := e^{i \vartheta} \bmm'.
\end{array}
\right.
&
\phantom{
\left\{
\begin{array}{l}
\\ \\ \\
\end{array}
\right.
}
\end{array}
\]
In general, \(a_1, a_2, a_3\) are complex-valued, and \(\vartheta\) is a real-valued scalar function of the coordinates.
However, to preserve the condition \(\mathring{\rho} = {\mathring{\bar{\rho}}}\), we restrict \(a_1, a_2, a_3\) to be real. Using the transformation formulae for the spin-coefficients given in Appendix~\ref{Appendix:LorentzTrans}, imposing \(\kappa'' = \kappa' = \kappa = 0\) reduces to solving the following equation for \(a_1\):
\begin{equation*}
\mathring{\kappa} + a_1 (\mathring{\sigma} + \mathring{\rho} + \mathring{\epsilon} - 2\,\operatorname{Re}(\mathring{\delta} a_1)) + a_1^2 (\mathring{\tau} + \mathring{\pi} + 2 \mathring{\beta} + 2 \mathring{\alpha} - \mathring{\Delta} a_1) + a_1^3 (\mathring{\mu} + \mathring{\lambda} + 2 \mathring{\gamma}) + a_1^4 \mathring{\nu} - \mathring{D} a_1 = 0.
\end{equation*}
    Once $a_1$ is determined, one can compute all the spin-coefficients associated to the frame $(\bml'',\bmn'',\bmm'',\bar{\bmm}'')$. With this information at hand, the next condition $\pi'=\pi=0$ is imposed by solving  the following equation for $a_2$:
    \begin{eqnarray*}
         \pi'' + 2 a_2 \epsilon'' + a_2{}^2 \kappa'' + D''a_2 =0.
    \end{eqnarray*}
    Having determined $a_2$ and computed all the spin-coefficients associated to the frame $(\bml',\bmn',\bmm',\bar{\bmm}')$ 
    we impose $\epsilon=0$ and $\tau=\bar{\alpha} + \beta$ which translates into solving the following equations for $a_3$ and $\vartheta$:
\begin{flalign}
  &  D' a_3 + a_3 e^{-i \vartheta} D' e^{i \vartheta} + 2 a_3\epsilon'=0,\\ &
2 \tau'
- \left(1 + e^{-2i\vartheta} \right) \delta' \ln a_3
+ \left(1-e^{-2i\vartheta} \right) e^{-i\vartheta} \delta' e^{i\vartheta}
- 2 \bar{\alpha}' - 2 e^{-2i\vartheta} \beta' =0.
\end{flalign}    
After completing this procedure, one effectively obtains the NP frame \( (\bml, \bmn, \bmm,\bar{\bmm} )\) along with their associated spin-coefficients and curvature components.  

\subsubsection*{Order of the expansion}

Although the preceding algorithm yields equations from which, in principle, exact expressions for \(a_1, a_2, a_3,\) and \(\vartheta\) could be derived, the complexity of these equations—combined with the detailed expressions for the original connection coefficients in Appendix~\ref{Apendix:Connection-Weyl-NP}—makes finding exact solutions intractable in practice.  
Consequently, we focus on obtaining asymptotic expansions for these parameters.  
The asymptotic solutions will be presented later; for now, since one of our main goals is to compute the NP constants and the BMS charges, it is essential to determine the order of expansion required to accurately evaluate these quantities.  
Since ---as shown in subsection~\ref{NP-coordinates}--- \(r = \mathfrak{r} + O(\mathfrak{r}^{-1})\), the order counting can be conveniently, and equivalently, carried out using the original radial coordinate \(r\).
As the NP constants depend on $\Psi_{0}$, we start by noticing that the composite transformation $\Lambda_3 \circ \Lambda_2 \circ \Lambda_ 1$ gives
\begin{equation*}
    \Psi_{0} = a_3{}^2 e^{2i \vartheta} (\mathring{\Psi}_0 + 4 a_1 \mathring{\Psi}_1 + 6 a_1{}^2 \mathring{\Psi}_2 + 4 a_1{}^3 \mathring{\Psi}_3 + a_1{}^4 \mathring{\Psi}_4).
\end{equation*}
From equations \eqref{original-Weyl-components}, we have that, for the $q$-metric, the Weyl scalars with respect to the tetrad $ (\mathring{\bml}, \mathring{\bmn}, \mathring{\bmm},\mathring{\bar{\bmm}}  )$  have the following fall-off:
\begin{eqnarray*}
    && \mathring{\Psi}_0 = \sum_{k=5}^{\infty} \frac{\mathring{\Psi}_{0}^{(k)}}{r^{k}} = \mathcal{O}(r^{-5}), \quad \mathring{\Psi}_1 = \sum_{k=5}^{\infty} \frac{\mathring{\Psi}_{1}^{(k)}}{r^{k}} = \mathcal{O}(r^{-5}), \quad \mathring{\Psi}_2= \sum_{k=3}^{\infty} \frac{\mathring{\Psi}_{2}^{(k)}}{r^{k}} = \mathcal{O}(r^{-3}), \\
    && \mathring{\Psi}_3= \sum_{k=5}^{\infty} \frac{\mathring{\Psi}_{3}^{(k)}}{r^{k}} = \mathcal{O}(r^{-5}), \quad \mathring{\Psi}_4= \sum_{k=5}^{\infty} \frac{\mathring{\Psi}_{4}^{(k)}}{r^{k}} = \mathcal{O}(r^{-5}).
\end{eqnarray*}
Therefore, the Weyl scalars $\mathring{\Psi}_0$ and $\mathring{\Psi}_2$ exhibit the decay rates expected from the standard peeling hierarchy, while $\mathring{\Psi}_1$, $\mathring{\Psi}_3$, and $\mathring{\Psi}_4$ decay faster than the generic peeling orders. This behaviour remains compatible with the peeling theorem and indicates a more special asymptotic structure than in the radiative case.
From the form of the above expansions, it is clear that,
\begin{equation}
    \Psi_0 = \sum_{k=5}^{\infty} \frac{\Psi_{0}^{(k)}}{r^{k}}.
    \label{Expansion-Psi0}
\end{equation}
Hence, the task is to write the expansion coefficient $\Psi_{0}^{(k)}$  in terms of 
%coefficients of $r^{-n}$ in the
analogous expansion coefficients for $a_1,a_3, \vartheta, \mathring{\Psi}_0,\mathring{\Psi}_1,\mathring{\Psi}_2,\mathring{\Psi}_3$ and $\mathring{\Psi}_4$. To do so, one introduces the following Ans\"{a}tze which are informed by the equations satisfied by $a_{1}, a_{2}, a_{3}$ and $\vartheta$
\begin{equation}
    a_1 = \sum_{k=2}^{\infty} \frac{a_1^{(k)}}{r^k}, \qquad a_2 = \sum_{k=2}^{\infty} \frac{a_2^{(k)}}{r^k}, \qquad a_3 = \sum_{k=0}^{\infty} \frac{a_3^{(k)}}{r^k}, \qquad \vartheta = \sum_{k=1}^{\infty} \frac{\vartheta^{(k)}}{r^k}. 
    \label{Ansatz-Lorentz-parameters}
\end{equation}
This calculation can be done systematically for all $k$, but for computing the NP constants, only $k=5$ and $k=6$ are required. A straightforward calculation renders
\begin{eqnarray*}
    && \Psi_{0}^{(5)} = a_{3}^{(0)}{}^2 \mathring{\Psi}_{0}^{(5)}, \\
    && \Psi_{0}^{(6)} = 2 a_{3}^{(0)} a_{3}^{(1)} \mathring{\Psi}_{0}^{(5)} + a_{3}^{(0)}{}^{2} \mathring{\Psi}_{0}^{(6)} + 2 i a_{3}^{(0)}{}^{2} \vartheta^{(1)} \mathring{\Psi}_{0}^{(5)}.
\end{eqnarray*}
On the other hand, for the calculation of the Bondi energy-momentum and the BMS charges we need to compute $\Psi_{2}$ accurately to order  $r^{-3}$. Hence, we also focus on this component and note that 
\begin{eqnarray*}
    && \Psi_{2} = \mathring{\Psi}_2 + 2 a_1 \mathring{\Psi}_3 + a_1{}^2 \mathring{\Psi}_4 + 2 a_2 (\mathring{\Psi}_1 + 3 a_1 \mathring{\Psi}_2 + 3 a_1{}^2 \mathring{\Psi}_3 + a_1{}^3 \mathring{\Psi}_4) \\
    && \phantom{\Psi_{2}}+ a_2{}^2 (\mathring{\Psi}_0 + 4 a_1 \mathring{\Psi}_1 + 6 a_1{}^2 \mathring{\Psi}_2 + 4 a_1{}^3 \mathring{\Psi}_3 + a_1{}^4 \mathring{\Psi}_4).
\end{eqnarray*}
Proceeding as before, one writes 
\begin{equation*}
    \Psi_2 = \sum_{k=3}^{\infty} \frac{\Psi_{2}^{(k)}}{r^{k}}.
\end{equation*}
With these elements, a direct calculation gives
\begin{equation*}
    \Psi_{2}^{(3)} = \mathring{\Psi}_{2}^{(3)}.
\end{equation*}
As a final remark, recall that the Weyl scalars for the original tetrad are given in equation \eqref{Original-tetrad-Weyl-components} in exact form; hence, they can be expanded to any desired order. Consequently, accurate expressions for $\Psi_{0}^{(5)}$ and $\Psi_{0}^{(6)}$ depend entirely on the asymptotic solutions for $a_{3}$ and $\vartheta$, to be explicitly calculated in the remainder of this subsection. 

\subsubsection*{Asymptotic solutions for the Lorentz transformations parameters}
The first step in the algorithm described above is to obtain an asymptotic solution for $a_{1}$. Using the Ansatz \eqref{Ansatz-Lorentz-parameters}, expanding the original spin-coefficients to order $r^{-4}$ and imposing the condition $\kappa''=0$, we find 
\begin{eqnarray*}
    && a_{1}^{(2)} = \frac{1}{2} m^2 q (2+q) \cos{\theta} \sin{\theta}, \\
    && a_{1}^{(3)} = - \frac{1}{8} m^3 q (-10+q+3q^2) \cos{\theta} \sin{\theta}.
\end{eqnarray*}
This solution implies that $\kappa''= \mathcal{O}(r^{-4})$ and it allows us to obtain asymptotic expressions for the transformed spin-coefficients $\tau'', \sigma'', \rho'', \ldots \text{etc}$. These expressions can then be used to fix the condition $\pi'=0$. Using the Ansatz \eqref{Ansatz-Lorentz-parameters}, one can show that the solution
\begin{eqnarray*}
    && a_{2}^{(2)} = - \frac{1}{2} m^2 q (2+q) \cos \theta \sin \theta,
\end{eqnarray*}
guarantees that $\pi'=\mathcal{O}(r^{-3})$. Using this solution, we can obtain an asymptotic expansion for the spin-coefficients $\tau', \sigma', \rho', \ldots \text{etc}$. Since $\kappa' = \kappa''$, this step leaves the order of $\kappa'$ unchanged. For the final step, we again use the Ansatz \eqref{Ansatz-Lorentz-parameters} and the explicit expressions of the transformed spin-coefficients to fix the conditions $\epsilon=0$ and $\tau-\bar{\alpha}-\beta=0$. Solving order by order, we obtain
\begin{eqnarray*}
    && a_{3}^{(0)} = 1, \quad a_{3}^{(1)} = m(1+q), \quad  a_{3}^{(2)} = \frac{1}{2} m^2 (3+4q+q^2),  \\
    && \vartheta^{(1)} = 0, \quad  \vartheta^{(2)} = 0.
\end{eqnarray*}
With this solution, the final expressions of the transformed  coefficients are given by
\begin{eqnarray*}
    && \kappa = \mathcal{O}(r^{-4}), \qquad \tau= \mathcal{O}(r^{-4}), \qquad \sigma = \mathcal{O}(r^{-4}), \\
    && \rho = - \frac{1}{\sqrt{2}r} + \frac{m q}{\sqrt{2}r^2} - \frac{m^2 q (-2 + 
   3 q + (2 + q) \cos(2 \theta))}{4 \sqrt{2}r^3} + \mathcal{O}(r^{-4}), \\
    && \epsilon = \mathcal{O}(r^{-4}), \qquad \gamma= \mathcal{O}(r^{-4}), \\
    && \beta = \frac{\cot{\theta}}{2 \sqrt{2} r} - \frac{m q \cot \theta}{2 \sqrt{2}r^2} + \frac{m^2 q (-6+q+(2+q)\cos (2 \theta)) \cot \theta}{8 \sqrt{2} r^3} + \mathcal{O}(r^{-4}), \\
    && \alpha = - \frac{\cot{\theta}}{2 \sqrt{2} r} + \frac{m q \cot \theta}{2 \sqrt{2}r^2} - \frac{m^2 q (-6+q+(2+q)\cos (2 \theta)) \cot \theta}{8 \sqrt{2} r^3} + \mathcal{O}(r^{-4}), \\
    && \pi = \mathcal{O}(r^{-3}), \qquad \nu = \mathcal{O}(r^{-3}), \\
    && \mu = - \frac{1}{\sqrt{2}r} + \frac{m (2+3q)}{\sqrt{2} r^2} + \mathcal{O}(r^{-3}), \\
    && \lambda = \mathcal{O}(r^{-3}).
\end{eqnarray*}
The transformed Weyl scalars are then given by
\begin{subequations}
\begin{eqnarray}
    && \Psi_{0} = \frac{m^3 q (1+q) (2+q) \sin^2{\theta}}{2 r^5} + \frac{5 m^4 q (1+q) (2+q) \sin^2{\theta}}{2 r^6} + \mathcal{O}(r^{-7}), \\
    && \Psi_{1} = -\frac{ m^3 q (1+q) (2+q) \sin{(2 \theta)}}{r^5} + \frac{m^4 q (1+q) (2+q) (-43 + 37 q) \sin{(2\theta)}}{16 r^6} + \mathcal{O}(r^{-7}), \qquad \quad \\
    && \Psi_{2} = -\frac{ m (1+q)}{r^3} + \frac{3 m^2 q (1+q)}{r^4} + \frac{m^3 q (1+q) (10-19q + 3(2+q) \cos{(2 \theta)})}{4 r^5}  \\
    && \phantom{\Psi_{2}} - \frac{m^4 q (1+q) (-38 + (93 - 64 q) q + 3 (2+q) (-7+8q) \cos{(2 \theta)})}{12 r^6} + \mathcal{O}(r^{-7}), \nonumber \\
    && \Psi_{3} = \frac{m^3 q (1+q) (2+q) \sin{(2\theta)}}{r^{5}} + \mathcal{O}(r^{-6}), \\
    && \Psi_{4} = \frac{ m^3 q (1+q) (2+q) \sin^2{\theta}}{2 r^5} - \frac{m^4 q (1+q) (2+q) (-1+4q) \sin^2{\theta}}{2 r^6} + \mathcal{O}(r^{-7}).
\end{eqnarray}
    \label{Weyl-coefficients}
\end{subequations}
While the expressions above are written in terms of the original coordinates $(t,r,\theta,\phi)$, one can  rewrite them in NP coordinates $(u,\mathfrak{r},\theta\,\phi)$ using equations  \eqref{Explicit-form-spherical-r} and \eqref{asymptotic-u-solution}, as we do in the following subsection.

\subsection{The Zipoy-Voorhees spacetime in the NP-gauge}
\label{Subsec:NP-quantities-NP-coordinates}
In this subsection we summarise the results obtained in previous sections by writing the spin-coefficients and Weyl scalars for the $q$-metric spacetime in the NP-gauge.

\medskip

First, for the spin-coefficients, we have
\begin{subequations}
    \begin{eqnarray}
    && \kappa = \mathcal{O}(\mathfrak{r}^{-4}), \qquad \tau= \mathcal{O}(\mathfrak{r}^{-4}), \qquad \sigma = \mathcal{O}(\mathfrak{r}^{-4}), \label{sigma-NP-coordinates} \\
    && \rho = - \frac{1}{\sqrt{2}\mathfrak{r}} + \frac{m q}{\sqrt{2}\mathfrak{r}^2} - \frac{m^2 q^2}{\sqrt{2} \mathfrak{r}^3} + \mathcal{O}(\mathfrak{r}^{-4}), \\
    && \epsilon = \mathcal{O}(\mathfrak{r}^{-4}), \qquad \gamma= \mathcal{O}(\mathfrak{r}^{-4}), \\
    && \beta = \frac{\cot{\theta}}{2 \sqrt{2} \mathfrak{r}} - \frac{m q \cot \theta}{2 \sqrt{2} \mathfrak{r}^2} + \frac{m^2 (-2+q) q \cot \theta}{4 \sqrt{2} \mathfrak{r}^3} + \mathcal{O}(\mathfrak{r}^{-4}), \\
    && \alpha = - \frac{\cot{\theta}}{2 \sqrt{2} \mathfrak{r}} + \frac{m q \cot \theta}{2 \sqrt{2} \mathfrak{r}^2} - \frac{m^2 (-2+q) q \cot \theta}{4 \sqrt{2} \mathfrak{r}^3} + \mathcal{O}(\mathfrak{r}^{-4}), \\
    && \pi = \mathcal{O}(\mathfrak{r}^{-3}), \qquad \nu = \mathcal{O}(\mathfrak{r}^{-3}), \\
    && \mu = - \frac{1}{\sqrt{2}\mathfrak{r}} + \frac{m (2+3q)}{\sqrt{2} \mathfrak{r}^2} + \mathcal{O}(\mathfrak{r}^{-3}), \\
    && \lambda = \mathcal{O}(\mathfrak{r}^{-3}).
    \end{eqnarray}
    \label{Connection-coefficients-NP-coordinates}
\end{subequations}
whereas the Weyl scalars fall-off as
\begin{eqnarray*}
    && \Psi_{0} = \frac{\Psi_{0}^{(5)}}{\mathfrak{r}^{5}} + \frac{\Psi_{0}^{(6)}}{\mathfrak{r}^{6}} + \mathcal{O}(\mathfrak{r}^{-7}), \\
    && \Psi_{1} = \frac{\Psi_{1}^{(5)}}{\mathfrak{r}^{5}} + \frac{\Psi_{1}^{(6)}}{\mathfrak{r}^{6}} + \mathcal{O}(\mathfrak{r}^{-7}), \\
    && \Psi_{2} = \frac{\Psi_{2}^{(3)}}{\mathfrak{r}^{3}} + \frac{\Psi_{2}^{(4)}}{\mathfrak{r}^{4}}+ \frac{\hat{\Psi}_{2}^{(5)}}{\mathfrak{r}^{5}} + \frac{\hat{\Psi}_{2}^{(6)}}{\mathfrak{r}^{6}} + \mathcal{O}(\mathfrak{r}^{-7}) , \\
    && \Psi_{3} = \frac{\Psi_{3}^{(5)}}{\mathfrak{r}^{5}} + \frac{\Psi_{3}^{(6)}}{\mathfrak{r}^{6}} + \mathcal{O}(\mathfrak{r}^{-7}), \\
    && \Psi_{4} = \frac{\Psi_{4}^{(5)}}{\mathfrak{r}^{5}} + \frac{\Psi_{4}^{(6)}}{\mathfrak{r}^{6}} + \mathcal{O}(\mathfrak{r}^{-7}), \\
\end{eqnarray*}
where $\Psi_n^{(k)}$ refer to the coefficients of $\mathfrak{r}^{-k}$ in the expansion of $\Psi_{n}$ as given in equation \eqref{Weyl-coefficients} by formally replacing $r$ by $\mathfrak{r}$. The coefficients $\hat{\Psi}_2^{(5)}$ and $\hat{\Psi}_2^{(6)}$ differ from the previous expansion and they are given by
\begin{subequations}
    \begin{eqnarray}
        && \hat{\Psi}_{2}^{(5)} = \frac{1}{2} m^3 q (1 + q) (2 - 11 q + 3 (2 + q) \cos(2 \theta)), \\
        && \hat{\Psi}_{2}^{(6)} = - \frac{5}{6} m^4 q (1 + q) (-2 + 3 q - 10 q^2 + (-6 + 9 q + 6 q^2) \cos(2 \theta)).
    \end{eqnarray}
\end{subequations}
These expressions allow us to compute explicitly the NP constants for the $q$-metric and other asymptotic conserved quantities. 
\section{Asymptotic charges for the Zipoy-Voorhees spacetime}
\label{Section:Asymptotic-conserved-quantities}
The asymptotic analysis of the $q$-metric carried out in the previous sections indicates that the deviation from spherical symmetry encoded by the parameter $q$ should be reflected in the structure of asymptotic quantities at null infinity. In particular, one expects asymptotic charges to capture signatures of the quadrupolar deformation parametrised by $q$. The aim of this section is to make this connection explicit by computing the Bondi--Sachs energy--momentum, the Newman--Penrose constants, and the BMS charges of the Zipoy-Voorhees spacetime.
%conserved quantities defined in the Newman-Penrose and the Bondi-Sachs formalisms. 
\subsubsection*{Bondi-Sachs energy-momentum}

The Bondi-Sachs energy-momentum as defined in  \cite{NewPen68}, is encoded through the following vector
\begin{equation*}
    \bm p = (P_{0,0}, \; 6^{-1/2}(P_{1,1}-P_{1,-1}), \; i 6^{-1/2}(P_{1,1}+P_{1,-1}), \; 3^{1/2}P_{1,0}),
\end{equation*}
in which $P_{n,m}$ is defined on a cut $\mathcal{C}$ of future null infinity $\mathscr{I}^{+}$ by
\begin{equation*}
    P_{n,m}:=-\frac{1}{\sqrt{32\pi}} \oint_{\mathcal{C}} {}_{0}{Y}_{n,m}(\sigma^{(2)} \dot{\bar{\sigma}}^{(2)} +\Psi_{2}^{(3)})  \bmd \mu,
\end{equation*}
where $\sigma^{(2)}$ refers to the leading term in $\sigma$ (the coefficient of the $\mathfrak{r}^{-2}$ term in the expansion), the dot notation refers to the derivative with respect to the retarded time $u$ so $\dot{\bar{\sigma}} := \bmpartial_{u} \bar{\sigma}$ and $\bmd \mu := \sin \theta \bmd \theta \bmd \phi$ is the standard volume element on $\mathbb{S}^{2}$. The Bondi mass is then given by 
$$
M_{B}:= \sqrt{P_{0,0}^2 - \tfrac{1}{3}P^{2}_{1,0} + \tfrac{2}{3}P_{1,1}P_{1,-1}}.
$$
The Bondi-Sachs energy-momentum loss formula is, in this notation, given by
\begin{equation*}
    \dot{P}_{n,m} = -\mathcal{F}_{n,m},
\end{equation*}
where
\begin{equation*}
    \mathcal{F}_{n,m}:= \oint_{\mathcal{C}} {}_{0}{Y}_{n,m}|\dot{\sigma}^{(2)}|^2   \bmd \mu.
\end{equation*}
The presence of gravitational radiation can be detected by the leading term in the shear $\sigma$ and naturally if $\dot{\sigma}^{(2)}=0$, there are no gravitational waves and therefore there is no loss of energy-mass through gravitational radiation and hence $Q_{n,m}=0$. 
Direct inspection of equation \eqref{sigma-NP-coordinates} shows that, as expected for a stationary spacetime,  $\sigma^{(2)}=0$ and hence the Bondi-Sachs energy-momentum is conserved and given explicitly by
\begin{equation*}
    \bm p = \Big(\frac{m (1+q)}{2 \sqrt{2}}, 0, 0, 0\Big).
\end{equation*}
Also notice that the Bondi mass is given by
\begin{equation*}
    M_{B} = \frac{m (1+q)}{2 \sqrt{2}}.
\end{equation*}
Naturally, in this case, the Bondi mass coincides with the ADM mass $M_{ADM}=m(1+q)$ computed in \cite{BosGasGutQueTor16} up-to a conventional normalisation factor.

\subsubsection*{The NP constants}
The NP constants are a set of quantities which, in contrast with the BMS charges, are absolutely conserved at any cut of null infinity, even in the presence of gravitational radiation ---see  \cite{NewPen68}.
For linear theories such as Maxwell's equations on a Minkowski spacetime background, there is an infinite number of these conserved quantities while for the full non-linear theory of General Relativity there are only 10. These are encoded in 5 complex quantities $G_{m}$ defined on a cut $\mathcal{C}$ of future null infinity $\mathscr{I}^+$ as follows:
\begin{equation*}
    G_{m} := \oint_{\mathcal{C}} {}_{2}\bar{Y}_{2,m} \Psi_{0}^{(6)}  \bmd \mu, \qquad m \in \{-2,-1,0,1,2 \}.
\end{equation*}
A direct calculation using the expression for $\Psi_{0}^{(6)}$ given in equation \eqref{Weyl-coefficients} renders
\begin{equation*}
    G_{-2} =0, \quad G_{-1}=0, \quad G_{0} = 2 m^4 \sqrt{\frac{10\pi}{3}} q (1 + q) (2 + q) , \quad G_{1} =0, \quad G_{2}=0.
\end{equation*}
Notice that, as expected, for $q=0$ and $q=-1$, all the Newman-Penrose constants vanish as these cases correspond to the Schwarzschild and Minkoski spacetimes, respectively.
To put this calculation in a more general context, notice that the form of $G_0$ is consistent with
the general expression given in \cite{NewPen68} valid for axisymmetric spacetimes:
\[
G_0 = 8 \sqrt{30\pi} (2D^2-MQ),
\]
where $M$, $D$ and $Q$ are the mass and complex dipole and quadrupole moments, respectively. In the case of the Kerr spacetime for instance one has that $D=iMa$ and $Q=-2Ma^2$ and hence $G_0$ vanishes because $2D^2$  coincides with $MQ$  ---see for instance \cite{Bac09}. Thus, it is clear that any axisymmetric spacetime with vanishing dipole moment will have non-trivial $G_0$ as long as its quadrupole moment is non-zero. Thus, this reinforces, at a conceptual level, that the Zipoy-Voorhees spacetime serves as a counter-example to the conjecture that \emph{the algebraically special condition in the theorem proved in \cite{WuSha07} can be relaxed to include  ``asymptotically algebraically special'' spacetimes} as argued in \cite{ZhaXiaGao09}. Therefore, these conditions are not enough to ensure that the NP constants vanish, and additional assumptions are implicitly made in the proof given in \cite{ZhaXiaGao09}. Finally, we comment that although arguably, one could have computed the NP constants for the Zipoy-Vorhees spacetime leveraging on the results of \cite{Bac09}, in practice one would have to re-express the metric and tetrad in the appropriate coordinates and gauge, hence we have opted to compute the NP constants directly from their original definition \cite{NewPen68} by recasting the Zipoy-Vorhees spacetime in the NP-gauge ---a gauge that is relevant for the asymptotic analysis of spacetimes beyond the specific calculation of the NP constants.
 
\subsubsection*{BMS-supertranslation charges at null infinity}
It is possible to compute the BMS-supertranslation charges associated with the $q$-metric using the expression in terms of NP-frame found in \cite{BarnLamb11}. For any $f \in C^{\infty}(\mathbb{S}^2)$, the associated BMS-supertranslation charge, defined on a cut $\mathcal{C}$ of future null infinity $\mathscr{I}^{+}$, is given by
\begin{equation*}
    Q_{f} := -\frac{1}{2 \sqrt{\pi}} \oint_{\mathcal{C}} f \; \text{Re} \left[ \Psi_{2}^{(3)}+ \sigma^{(2)} \dot{\bar{\sigma}}^{(2)} \right] \bmd \mu.
\end{equation*}
For simplicity, one can always choose $f = Y_{l,m}$ so that for each $Y_{l,m}$, the associated BMS charge is given by
\begin{equation}
    Q_{l,m} := -\frac{1}{2 \sqrt{\pi}} \oint_{\mathcal{C}} Y_{l,m} \; \text{Re} \left[ \Psi_{2}^{(3)}+ \sigma^{(2)} \dot{\bar{\sigma}}^{(2)} \right] \bmd \mu,  \qquad \text{on } \mathscr{I}^{+}.
    \label{BMS-charge-l-m}
\end{equation}
Using the expressions obtained in section \ref{Subsec:NP-quantities-NP-coordinates}, we can show that 
\begin{equation*}
    \text{Re} \left[ \Psi_{2}^{(3)}+ \sigma^{(2)} \dot{\bar{\sigma}}^{(2)} \right] = - m (1+q).
\end{equation*}
Substituting into \eqref{BMS-charge-l-m}, we get
\begin{equation*} Q_{l,m} = 
    \begin{cases}
         m (1+q), \qquad & \text{for } l =0, m=0, \\
         0, \qquad & \text{for } l \neq 0,  m \neq 0.
    \end{cases}
\end{equation*}
As expected, the $l=0, m=0$ charge correspond to the ADM mass. We also note that all \emph{supertranslation} charges are vanishing. Analogous expressions for $Q_{l,m}$ can be obtained at $\mathscr{I}^{-}$.

%\bigskip

\section{Conclusions}
\label{Sec:Conclusions}

The NP constants (Newman--Penrose) and BMS charges (Bondi-Metzner-Sachs) are asymptotic quantities defined as integrals of the Weyl curvature over cuts of future null infinity. Although both sets of quantities can be computed in a similar manner, they exhibit distinct physical and mathematical properties. The NP constants form a set of ten absolutely conserved quantities, remaining constant even in the presence of gravitational radiation. In contrast, the BMS charges constitute an infinite set of quantities associated with asymptotic symmetries at null infinity, and are sensitive to the flux of radiation through null infinity.
%%%%
Despite the relatively straightforward definitions of the NP constants and BMS charges, explicit calculations of these quantities are scarce in the literature. This is in part because most exact solutions are not expressed in the gauge of the classical theory of asymptotics of Newman and Penrose \cite{NewPen62}, in which, for example, the NP constants are naturally defined ---see \cite{NewPen68}. Notably, it has been shown that the NP constants of the Kerr spacetime vanish identically. Furthermore, a theorem proved in \cite{WuSha07} establishes that the NP constants also vanish for all stationary, asymptotically flat, algebraically special solutions to the vacuum Einstein field equations.

\medskip

Motivated by these results, this paper presents a computation of the NP constants and BMS charges for a simple yet algebraically general vacuum solution to the Einstein equations: the Zipoy--Voorhees spacetime. This spacetime is a static, axisymmetric solution that depends on two parameters, \( m \) and \( q \), which are related to the mass and quadrupole moment, respectively. For \( q = 0 \), the solution reduces to the Schwarzschild metric, allowing \( q \) to be interpreted as a deformation parameter and hence the Zipoy-Voorhees solution is sometimes referred to as the $q$-metric. 
%%%
Due to its relative simplicity, this solution has been widely used in studies exploring possible astrophysical signatures of naked singularities ---see for instance \cite{BosGasGutQueTor16, BosKonEtAl21, SerKuaQueEtAl24, DanUteQueUra25}. In this work, however, we focus on the asymptotic structure of this spacetime by expressing it in the NP-gauge. Although only formal asymptotic expansions are obtained, they are sufficient to compute the NP constants, the Bondi-Sachs energy-momentum and BMS charges explicitly. 
We find, in particular, that the NP constants of the Zipoy--Voorhees spacetime are nonzero. This result highlights the importance of the algebraically special condition in the theorem proven in \cite{WuSha07}.
Furthermore, it is shown that either the algebraically special condition cannot be relaxed to ``asymptotically algebraically special'' as argued in \cite{ZhaXiaGao09} or there are extra implicit assumptions in that result that should exclude the case of the Zipoy-Voorhees spacetime ---and probably other spacetimes with non-trivial quadrupole moments such as those discussed in \cite{FruQueSan18}.
%%%

\medskip

Beyond their standard interpretation as conserved quantities at null infinity, the NP-constants fit naturally into the wider hierarchy of asymptotic symmetries and their charges, particularly through their connection to higher-spin symmetries and their associated charges; see for instance \cite{FreiPranRacl21,CrestFrei24}. As shown in these works, an expansion of $\Psi_{0}$ of the form \eqref{Expansion-Psi0} allows each coefficient $\Psi^{(k)}_{0}, k\geq5$ to be decomposed into a global and a local component. The global component further splits into harmonics with $l \in [2, k-4]$, while the local component contains the $l \geq k-3$ modes. The former yields the globally conserved quantities, i.e., the NP-constants, whereas the latter carries the higher-spin asymptotic charges, which satisfy certain evolution equations on null infinity. As already established in the original work of Newman and Penrose \cite{NewPen68}, linearised gravity admits an infinite family of such global components, and therefore an infinite number of NP-constants. This arises because, in the linear theory, the different harmonic modes do not mix: both the global and local components satisfy separate evolution equations, and the structure of these equations guarantees the conservation of the global part. In the non-linear setting, the mixing of harmonics leads to coupled evolution equations for the global and local components, and this coupling only disappears at the level of $\Psi^{(6)}_{0}$. Therefore, the analysis here highlights how sensitive the global component of $\Psi^{(6)}_{0}$ is to the assumptions placed on the background spacetime: even for asymptotically flat and static spacetimes, the asymptotically algebraically special condition seems to be insufficient to ensure the vanishing of this component. 
%%%%%%%%%%%%%%%%%%%%%%%%%%%

\medskip

\subsubsection*{Acknowledgements}

We are grateful to J. A. Valiente Kroon, Nicolas Cresto, Laurent Friedel and J. L. Jaramillo for their helpful comments on this manuscript. Calculations in this project used the computer algebra system Mathematica with the package xAct \cite{xAct}.
EG held an FCT investigator grant (2020.03845.CEECIND) during the early stages of this research project. 
MM gratefully acknowledges support from Perimeter Institute for Theoretical Physics through the Fields-AIMS-Perimeter fellowship. Research at Perimeter Institute is supported by the Government of Canada through the Department of Innovation, Science and Economic Development and by the Province of Ontario through the Ministry of Colleges and Universities. This work was supported by the Simons Collaboration on Celestial Holography.

%\newpage
\appendix
\setcounter{equation}{0}  % reset counter 
\renewcommand{\theequation}{\thesection.\arabic{equation}}
%%%%%%%%%%%%%%%%%%%%%%%%%%%%%%%%%%%%%%%%%%%%%%%%

\section{Spin connections and Weyl Components in NP-gauge}
\label{Apendix:Connection-Weyl-NP}
Following the notations of \cite{Stewart91} and starting with the null tetrad \eqref{Original-null-tetrad}, a direct calculation gives the following spin-coefficients:
\begin{subequations}\label{eq:RingSpinCoef}
\begin{flalign}
    & \mathring{\kappa}  = -\frac{F^{q/2} G^{q (2 + q)/2} m^2 q (2 + q) \cos\theta  \sin\theta }{2 \sqrt{2} r (r (-2 m + r) + m^2 \sin^2\theta )} ,\\
   &\mathring{\tau}  = \frac{F^{q/2} G^{q (2 + q)/2} m^2 q (2 + q) \cos\theta  \sin\theta }{2 \sqrt{2} r (r (-2 m + r) + m^2 \sin^2\theta )},\\
   &\mathring{\sigma} = \frac{F^{(q-1)/2} G^{q (2 + q)/2} m^2 q (2 + q) (m  - r) \sin^2\theta }{2 \sqrt{2} r^2 (r (-2 m + r) + m^2 \sin^2\theta )},\\
   &\mathring{\rho}  = \frac{F^{(q-1)/2} G^{q (2 + q)/2} (2 (r-2 m ) (m (2 + q)  - r) r + m^2 (m (2 + q)^2  - (2 + q (2 + q)) r) \sin^2\theta )}{2 \sqrt{2} r^2 (r (-2 m + r) + m^2 \sin^2\theta )},\\
   & \mathring{\epsilon}  = \frac{F^{(q-1)/2} G^{q (2 + q)/2} m (1 + q)}{2 \sqrt{2} r^2},\\
   & \mathring{\gamma}  = \frac{F^{(q-1)/2} G^{q (2 + q)/2} m (1 + q)}{2 \sqrt{2} r^2},\\
   & \mathring{\beta}  = \frac{F^{q/2} G^{q (2 + q)/2} \cot\theta }{2 \sqrt{2} r},\\
   & \mathring{\alpha}  = - \frac{F^{q/2} G^{q (2 + q)/2} \cot\theta }{2 \sqrt{2} r},\\
   & \mathring{\pi}  = - \frac{F^{q/2} G^{ q (2 + q)/2} m^2 q (2 + q) \cos\theta  \sin\theta }{2 \sqrt{2} r (r (-2 m + r) + m^2 \sin^2\theta )},\\
   & \mathring{\nu}  =  \frac{F^{q/2} G^{ q (2 + q)/2} m^2 q (2 + q) \cos\theta  \sin\theta }{2 \sqrt{2} r (r (-2 m + r) + m^2 \sin^2\theta )},\\
   & \mathring{\mu} = \frac{F^{(q-1)/2} G^{q (2 + q)/2} (2 (r -2 m  ) (m (2 + q)  - r) r + m^2 (m (2 + q)^2  - (2 + q (2 + q)) r) \sin^2\theta )}{2 \sqrt{2} r^2 (r (-2 m + r) + m^2 \sin^2\theta )},\\
  & \mathring{ \lambda}  =  \frac{F^{(q-1)/2} G^{q (2 + q)/2} m^2 q (2 + q) (m  - r) \sin^2\theta }{2 \sqrt{2} r^2 (r (-2 m + r) + m^2 \sin^2\theta )},
\end{flalign}
\end{subequations}
where $F$ and $G$ are defined in equation \eqref{Definition-F-G}. The components of the Weyl spinor associated with the original null tetrad $( \mathring{\bml}, \mathring{\bmn}, \mathring{\bmm}, \mathring{\bar{\bmm}})$ are given by \cite{NewPen62}
\begin{subequations}
    \begin{eqnarray}
    && \mathring{\Psi} _{0}{} = \frac{F^q G^{q (2 + q)} m^3 q (1 + q) (2 + q) (m  - r) \sin^2\theta }{2 (2 m  - r) r^3 (r (-2 m + r) + m^2 \sin^2\theta )},\\
    && \mathring{\Psi} _{1}{} = - \frac{F^{q-1/2} G^{q (2 + q)} m^3 q (1 + q) (2 + q) \cos\theta  \sin\theta }{2 r^3 (r (-2 m + r) + m^2 \sin^2\theta )},\\
    && \mathring{\Psi} _{2}{} = \frac{F^q G^{q (2 + q)} m(1+q)}{2 (2 m  - r) r^3 (m^2 + \csc^2\theta  r (-2 m + r))}, \nonumber \\
    && \phantom{\mathring{\Psi} _{2}{}} \times(2 \csc^2\theta  (2 m  - r) (m (2 + q)  - r) r + m^2 (- m (2 + q)^2 + (2 + q (2 + q)) r)), \quad  \\
    && \mathring{\Psi} _{3}{} = \frac{F^{q-1/2} G^{q (2 + q)} m^3 q (1 + q) (2 + q) \cos\theta  \sin\theta }{2 r^3 (r (-2 m + r) + m^2 \sin^2\theta )},\\
    && \mathring{\Psi} _{4}{} = \frac{F^q G^{q (2 + q)} m^3 q (1 + q) (2 + q) (m  - r) \sin^2\theta }{2 (2 m  - r) r^3 (r (-2 m + r) + m^2 \sin^2\theta )}.
    \end{eqnarray}
    \label{Original-tetrad-Weyl-components}
\end{subequations}

\section{The \texorpdfstring{$q$}{q}-metric in NP coordinates}
\label{Appendix:CoordinatesExpansions}

As discussed in subsection \ref{NP-coordinates}, the first step to obtain NP coordinates $(u,\mathfrak{r},\theta,\phi)$ is to solve the eikonal equation to obtain the retarded time $u=u(t,r,\theta,\phi)$, albeit in asymptotic form.
Using equation \eqref{asymptotic-u-solution}, a direct calculation shows that the $q$-metric components in the chart $(u,r,\theta,\phi)$ satisfy the following expansions:
\begin{eqnarray*}
    && g_{uu} = 1 - \frac{2 m (1+q)}{r} + \frac{2 m^2 q (1+q)}{r^2} - \frac{4 m^3 q (q^2-1)}{3r^3} + \mathcal{O}(r^{-4}), \\
    && g_{ur} = g_{ru} = 1 - \frac{m^2 q (2+q) \sin^2\theta}{2r^2} - \frac{m^3 q (2+q) \sin^{2}\theta}{r^3} + \mathcal{O}(r^{-4}), \\
    && g_{u \theta} = g_{\theta u} = \sum_{n=1}^{3} \frac{g_{u \theta}^{(n)}}{r^n} + \mathcal{O}(r^{-4}), \\
    && g_{r \theta} = g_{\theta r} = \sum_{n=1}^{3} \frac{g_{r \theta}^{(n)}}{r^n}+ \mathcal{O}(r^{-4}), \\
    && g_{\theta \theta} = -r^2 - 2 m r q + m^2 q (-2 (1+q)+(2+q) \sin^2\theta) +\sum_{n=1}^{3} \frac{g_{\theta \theta}^{(n)}}{r^n}+ \mathcal{O}(r^{-4}), \\
    && g_{\phi \phi} = - r^2 \sin^2\theta - 2 m r q \sin^2\theta - 2 m^2 q (1+q) \sin^2\theta+ \sum_{n=1}^{3} \frac{g_{\phi \phi}^{(n)}}{r^n}+ \mathcal{O}(r^{-4}), \\
\end{eqnarray*}
where all other components are $\mathcal{O}(r^{-4})$ or higher order. Here, the coefficients $g_{u \theta}^{(n)}, g_{r \theta}^{(n)},g_{r \phi}^{(n)}, g_{\theta \theta}^{(n)}$ and $g_{\phi \phi}^{(n)}$ are given by
\begin{eqnarray*}
    && g_{u \theta}^{(1)} =  m^2 q (2+q) \cos\theta \sin\theta, \\
    && g_{u \theta}^{(2)} = - m^3 q^2 (q+2) \sin\theta \cos\theta, \\ 
    && g_{u \theta}^{(3)} = \frac{1}{24} m^4 q (2+q) (-2-10 q+7 q^2+(2+10 q+5 q^2) \cos(2\theta)) \sin(2 \theta), \\
    && g_{r \theta}^{(1)} = m^2 q (2+q) \cos\theta \sin\theta, \\
    && g_{r \theta}^{(2)} = \frac{1}{2} m^3 q (2 + q)^2 \sin(2 \theta), \\ 
    && g_{r \theta}^{(3)} = \frac{1}{12} m^4 q (2+q) (23+16 q+2 q^2+(1+8 q+4 q^2) \cos(2 \theta)) \sin(2 \theta), \\
    && g_{\theta \theta}^{(1)} = - \frac{1}{3} m^3 q (2+3 q+q^2) (1+3 \cos(2 \theta)), \\
    && g_{\theta \theta}^{(2)} = - \frac{1}{48} m^4 q (2+q) (9-10 q-13 q^2+12 (7+10 q+3 q^2) \cos(2 \theta)+3 (1+6 q+3 q^2) \cos(4 \theta)), \\ 
    && g_{\theta \theta}^{(3)} = - \frac{1}{120} m^5 q (2+q) (-6-107 q-74 q^2-3 q^3+20 (18+29 q+12 q^2+q^3) \cos(2 \theta) \\
    && \phantom{g_{\theta \theta}^{(3)} =}+15 (2+9 q+6 q^2+q^3) \cos(4 \theta)), \\
    && g_{\phi \phi}^{(1)} = - \frac{4}{3} m^3 q (2+3 q+q^2)\sin^2\theta, \\
    && g_{\phi \phi}^{(2)} = - \frac{2}{3} m^4 q (6+11 q+6 q^2+q^3) \sin^2\theta, \\ 
    && g_{\phi \phi}^{(3)} = - \frac{4}{15} m^5 q (24+50 q+35 q^2+10 q^3+q^4) \sin^2\theta.
\end{eqnarray*}

Notice that the coordinates $(u,r,\theta,\phi)$ do not yet correspond to the NP coordinates. The key difference is in the definition of the ``radial" coordinate. In the theory of asymptotics, there are different choices for this. For example, in the Bondi-Sachs formalism ---see \cite{Mad16} for a modern review--- the areal radius is chosen. On the other hand, in the NP theory of asymptotics, one instead chooses an affine parameter $\mathfrak{r}$ along null geodesics ---see \cite{NewPen62}. For a discussion of the relation between these two historical choices, see \cite{Val98}. In the case of the NP-gauge, choosing $\mathfrak{r}$ as a coordinate, ensures that  $\bmg^\sharp$ has the following form
\begin{equation*}
    \bmg^{\sharp} = \begin{pmatrix}
0 & 1 & 0 & 0\\
1 & g^{\mathfrak{r}\mathfrak{r}} & g^{\mathfrak{r}\theta} & g^{\mathfrak{r}\phi}\\
0 & g^{\theta\mathfrak{r}} & g^{\theta\theta} & g^{\theta\phi}\\
0 & g^{\phi\mathfrak{r}} & g^{\phi\theta} & g^{\phi\phi}\\
\end{pmatrix},
\end{equation*}
Using expression \eqref{Explicit-form-spherical-r}, a calculation shows that the $q$-metric components in the coordinates $(u,\mathfrak{r},\theta,\phi)$ satisfy the following expansions:
\begin{subequations}
    \begin{eqnarray}
    && g^{\mathfrak{r}\mathfrak{r}}  = - 1 + \frac{2 m (1+q)}{\mathfrak{r}} - \frac{2m^2 q (1+q)}{\mathfrak{r}^2} + \mathcal{O}(\mathfrak{r}^{-3}), \\
    && g^{\theta \theta } = - \frac{1}{\mathfrak{r}^2} + \mathcal{O}(\mathfrak{r}^{-3}), \\
    && g^{\phi \phi} = -\frac{\csc^2{\theta}}{\mathfrak{r}^2} + \mathcal{O}(\mathfrak{r}^{-3}),
\end{eqnarray}
\label{eq:MetricInNPcoords}
\end{subequations}
with all other terms $\mathcal{O}(\mathfrak{r}^{-3})$ or higher. More detailed and higher order expansions can be algorithmically computed using the method described in the main text.

\section{Lorentz transformations}
\label{Appendix:LorentzTrans}

The transformation formulae for the spin-coefficients and the Weyl curvature associated to the transformations $\Lambda_1$, $\Lambda_2$ and $\Lambda_3$ can be found in \cite{ODonnell03} and are written here for the convenience of the reader. As customary in the NP-formalism, the directional derivatives along $\bml,\bmn$, $\bmm$, and $\bar{\bmm}$ will be denoted as $D$ $\Delta$, $\delta$, and $\bar{\delta}$, respectively. In the following expressions, each object will be appropriately decorated to indicate the tetrad with respect to which it is defined.

\subsection{Null rotation with \texorpdfstring{$n$}{n} fixed}
\begin{align}
    \label{Step-1-null-tetrad-transformation}
    \Lambda_1:   \qquad  \begin{cases}
     \mathring{\bml} \to \bml'' := \mathring{\bml} + a_{1} \mathring{\bmm} + a_1 \bar{\mathring{\bmm}} + a_1{}^2  \mathring{\bmn},\\
     \mathring{\bmn} \to \bmn'' := \mathring{\bmn}, \\     
     \mathring{\bmm} \to \bmm'' := \mathring{\bmm} + a_1 \mathring{\bmn},       \end{cases} 
\end{align}
where $a_1$ is, in general, a complex-valued function of the coordinates.
Under this transformation, the spin-coefficients transform as follows

\begin{eqnarray*}
    && \mathring{\kappa} \to \kappa'' := \mathring{\kappa} + a_1{}^2 \mathring{\tau} + a_{1} \mathring{\sigma} + a_1 \mathring{\rho} + a_1{}^2 \mathring{\pi} + a_1{}^4 \mathring{\nu} + a_1{}^3 \mathring{\mu} + a_1{}^3 \mathring{\lambda} + a_1 \mathring{\epsilon} + 2 a_1{}^3 \mathring{\gamma} \\
    && \phantom{\mathring{\kappa} \to \kappa }+ 2 a_1{}^2 \mathring{\beta} + 2 a_1{}^2 \mathring{\alpha} - \mathring{D} a_1 - a_1{}^2  \mathring{\Delta} a_1 - a_{1} \mathring{\delta} a_1 - a_1 \bar{\mathring{\delta}} a_1, \\
    && \mathring{\tau} \to \tau'' := \mathring{\tau} + 2 a_1 \mathring{\gamma} + a_1{}^2 \mathring{\nu} - \mathring{\Delta} a_1, \\
    && \mathring{\sigma} \to \sigma'' := \mathring{\sigma} + a_1 \mathring{\tau} + 2 a_1 \mathring{\beta} + 2 a_1{}^2 \mathring{\gamma} + a_1{}^2 \mathring{\mu} + a_1{}^3 \mathring{\nu} - \mathring{\delta} a_1 - a_1 \mathring{\Delta} a_1, \\
    && \mathring{\rho} \to \rho'' := \mathring{\rho} + a_{1} \mathring{\tau} + a_1{}^2 \mathring{\lambda} + a_1{}^3  \mathring{\nu} + 2 a_1 \mathring{\alpha} + 2 a_1{}^2 \mathring{\gamma} - \bar{\mathring{\delta}} a_1 - a_{1} \mathring{\Delta} a_1, \\
    && \mathring{\epsilon} \to \epsilon'' := \mathring{\epsilon} + a_1{}^2 \mathring{\gamma} + a_{1} \mathring{\beta} + a_1 \mathring{\alpha} + a_1 \mathring{\pi} + a_1{}^3 \mathring{\nu} + a_1{}^2 \mathring{\mu} + a_1{}^2 \mathring{\lambda}, \\
    && \mathring{\gamma} \to \gamma'' := \mathring{\gamma} + a_1 \mathring{\nu}, \\
    && \mathring{\beta} \to \beta'' := \mathring{\beta} + a_1 \mathring{\gamma} + a_1 \mathring{\mu} + a_1{}^2 \mathring{\nu}, \\
    && \mathring{\alpha} \to \alpha'' := \mathring{\alpha} + a_{1} \mathring{\tau} + a_1 \mathring{\lambda} + a_1{}^2 \mathring{\nu}, \\
    && \mathring{\pi} \to \pi'' := \mathring{\pi} + a_1 \mathring{\lambda} + a_{1} \mathring{\mu} + a_1{}^2 \mathring{\nu}, \\
    && \mathring{\nu} \to \nu'' := \mathring{\nu}, \\
    && \mathring{\mu} \to \mu'' := \mathring{\mu} + a_1 \mathring{\nu}, \\
    && \mathring{\lambda} \to \lambda'' := \mathring{\lambda} + a_{1} \mathring{\nu}, 
\end{eqnarray*}
and the Weyl scalars as
\begin{eqnarray*}
    && \mathring{\Psi}_{0} \to \Psi''_{0} := \mathring{\Psi}_0 + 4 a_1 \mathring{\Psi}_1 + 6 a_1{}^2 \mathring{\Psi}_2 + 4 a_1{}^3 \mathring{\Psi}_3 + a_1{}^4 \mathring{\Psi}_4, \\
    && \mathring{\Psi}_{1} \to \Psi''_{1} := \mathring{\Psi}_1 + 3 a_1 \mathring{\Psi}_2 + 3 a_1{}^2 \mathring{\Psi}_3 + a_1{}^3 \mathring{\Psi}_4, \\
    && \mathring{\Psi}_{2} \to \Psi''_{2} := \mathring{\Psi}_2 + 2 a_1 \mathring{\Psi}_3 + a_1{}^2 \mathring{\Psi}_4, \\
    && \mathring{\Psi}_{3} \to \Psi''_{3} := \mathring{\Psi}_3 + a_1 \mathring{\Psi}_4, \\
    && \mathring{\Psi}_{4} \to \Psi''_{4} := \mathring{\Psi}_4.
\end{eqnarray*}

\subsection{Null rotation with \texorpdfstring{$l$}{l} fixed}
\begin{align}
    \label{Step-2-null-tetrad-transformation}
    \Lambda_2: \qquad  \begin{cases}
     {\bml}'' \to \bml' := {\bml}'',\\
    \bmn'' \to \bmn' := \bmn''+ a_{2} \bmm'' +a_2 \bar{\bmm}'' + a_2{}^2 \bml'',  \\     
     \bmm'' \to \bmm' := \bmm'' + a_2 \bml'' ,     \end{cases} 
\end{align}
where $a_2$ is, in general, a complex-valued function of coordinates.
Under this transformation, the spin-coefficients transform as follows
\begin{eqnarray*}
    && \kappa'' \to \kappa' := \kappa'', \\
    && \tau'' \to \tau' := \tau'' + a_2 \sigma'' + a_2 \rho'' + a_2{}^2 \kappa'', \\
    && \sigma'' \to \sigma':= \sigma'' + a_2 \kappa'', \\
    && \rho'' \to \rho' := \rho'' + a_2 \kappa'', \\
    && \epsilon'' \to \epsilon' := \epsilon'' + a_2 \kappa'', \\
    && \gamma'' \to \gamma' := \gamma'' + a_2{}^2 \epsilon'' + a_2 \alpha'' + a_2 \beta'' + a_2 \tau'' + a_2{}^3 \kappa'' + a_2{}^2 \rho'' +a_2{}^2 \sigma'', \\
    && \beta'' \to \beta' := \beta'' + a_2 \epsilon'' + a_2 \sigma'' + a_2{}^2 \kappa'', \\
    && \alpha'' \to \alpha' := \alpha'' + a_2 \epsilon'' + a_2 \rho'' + a_2{}^2 \kappa'', \\
    && \pi'' \to \pi' := \pi'' + 2 a_2 \epsilon'' + a_2{}^2 \kappa'' + D''a_2, \\
    && \nu'' \to \nu' := \nu'' + a_2{}^2 \pi'' + a_2 \lambda'' + a_2 \mu'' + a_2{}^2 \tau'' + a_2{}^2 \kappa'' + a_2{}^3 \rho'' + a_2{}^3 \sigma'' + 2 a_2 \gamma''  \\
    && \phantom{\nu'' \to \nu' } + 2 a_2{}^3 \epsilon'' + 2 a_2 \alpha'' + 2 a_2{}^2 \beta'' + \Delta'' a_2 + a_2{}^2 D'' a_2 + a_2 \bar{\delta}'' a_2 + a_2 \delta'' a_2, \\
    && \mu'' \to \mu' := \mu'' + a_2 \pi'' + a_2{}^2 \sigma'' + a_2{}^3 \kappa'' + 2 a_2 \beta'' + 2 a_2{}^2 \epsilon'' + \delta'' a_2 + a_2 D'' a_2, \\
    && \lambda'' \to \lambda' := \lambda'' + a_2 \pi'' + 2 a_2 \alpha'' + 2 a_2{}^2 \epsilon'' + a_2{}^2 \rho'' + a_2{}^3 \kappa'' + \bar{\delta}'' a_2 + a_2 D'' a_2,
\end{eqnarray*}
and the Weyl scalars as
\begin{eqnarray*}
    && \Psi''_{0} \to \Psi'_{0} := \Psi''_{0}, \\
    && \Psi''_{1} \to \Psi'_{1} := \Psi''_{1} + a_2 \Psi''_{0}, \\ 
    && \Psi''_{2} \to \Psi'_{2} := \Psi''_{2} + 2 a_2 \Psi''_{1} + a_2{}^2 \Psi''_{0}, \\
    && \Psi''_{3} \to \Psi'_{3} := \Psi''_{3} + 3 a_{2} \Psi''_{2} + 3 a_2{}^2 \Psi''_{1} + a_2{}^3 \Psi''_{0}, \\
    && \Psi''_{4} \to \Psi'_{4} := \Psi''_{4} + 4 a_2 \Psi''_{3} + 6 a_2{}^2 \Psi''_{2} + 4 a_2{}^3 \Psi''_{1} + a_2{}^4 \Psi''_{0},
\end{eqnarray*}

\subsection{Spin-boost}
\begin{align}
    \label{Step-3-null-tetrad-transformation}
    \Lambda_3: \qquad  \begin{cases}
     \bml' \to \bml := a_3 \bml',\\
    \bmn' \to \bmn := \frac{1}{a_3} \bmn',  \\     
     \bmm' \to \bmm := e^{i \vartheta} \bmm',  \end{cases} 
\end{align}
where $a_3$ and $\vartheta$ are, respectively, a complex-valued and a real-valued function of the coordinates.
In this case, the spin-coefficients transform as,

\begin{eqnarray*}
    && \kappa' \to \kappa := a_3{}^2 e^{i \vartheta} \kappa', \\
    && \tau' \to \tau := e^{i\vartheta} \tau', \\
    && \sigma' \to \sigma:= a_3 e^{2 i \vartheta} \sigma', \\
    && \rho' \to \rho := a_3 \rho ,\\
    && \epsilon' \to \epsilon := \frac{a_3}{2} \left( \frac{1}{a_3} D' a_3 + e^{-i \vartheta} D' e^{i \vartheta} + 2 \epsilon' \right), \\
    && \gamma' \to \gamma := \frac{1}{2 a_3} \left( \frac{1}{a_3} \Delta' a_3 + e^{-i \vartheta} \Delta e^{i \vartheta} + 2 \gamma' \right), \\
    && \beta' \to \beta := \frac{1}{2} e^{-i \vartheta} \left( \frac{1}{a_3} \delta' a_3 + e^{-i \vartheta} \delta' e^{i \vartheta} + 2 \beta' \right), \\
    && \alpha' \to \alpha := \frac{1}{2} e^{-i \vartheta} \left( \frac{1}{a_3} \bar{\delta}' a_3 + e^{-i \vartheta} \bar{\delta}' e^{i \vartheta} + 2 \alpha' \right), \\
    && \pi' \to \pi := e^{-i \vartheta} \pi', \\
    && \nu' \to \nu := \frac{1}{a_3} e^{-i \vartheta} \nu', \\
    && \mu' \to \mu := \frac{1}{a_3} \mu', \\
    && \lambda' \to \lambda := \frac{1}{a_3} e^{-2 i \vartheta} \lambda',
\end{eqnarray*}
and the Weyl scalars as,
\begin{eqnarray*}
    && \Psi'_{0} \to \Psi_{0} := a_3{}^2 e^{2 i \vartheta} \Psi'_{0}, \\
    && \Psi'_{1} \to \Psi_{1} := a_3 e^{i \vartheta} \Psi'_{1}, \\
    && \Psi'_{2} \to \Psi_{2} := \Psi'_{2}, \\
    && \Psi'_{3} \to \Psi_{3} := a^{-1} e^{-i \vartheta} \Psi'_{3}, \\
    && \Psi'_{4} \to \Psi_{4} := a^{-4} e^{-2 i \vartheta} \Psi'_{4}.
\end{eqnarray*}
\section{Initial data and asymptotic flatness}
\label{Ap:AsymptEuclidean}
To show that the Zipoy–Voorhees spacetime is asymptotically Euclidean and regular as per the definition in \cite{Geroch70}, we first start by recalling this definition for a general three-dimensional Riemannian manifold
\begin{definition}
    \emph{A three-dimensional Riemannian manifold $(\mathcal{S},\bmh)$ is \emph{asymptotically Euclidean and regular} if there exists a three-dimensional, orientable, compact manifold $(\bar{\mathcal{S}},\bar{\bmh})$ with points $i_k \in \mathcal{S}, k = 1, \dotso, N$ with $N$ some integer, a function $\bar{\Omega} \in C^2$ and a diffeomorphism $\varphi: \bar{\mathcal{S}} \setminus \{i_1, \dotso,i_N \} \rightarrow \mathcal{S}$ such that
    \begin{enumerate}
        \item $\bar{\Omega}(i_k)=0, \bmd \bar{\Omega}(i_k) =0$ and $\textbf{Hess } \bar{\Omega}(i_k) = -2 \bar{\bmh}(i_k)$, for all $i_k \in \{i_1, \dotso,i_N \}$,
        \item $\bar{\Omega} > 0$ on $\bar{\mathcal{S}} \setminus \{i_1, \dotso,i_N \}$, and
        \item $\bar{\bmh} = \bar{\Omega}^2 \varphi^{*} \bmh$ on $\bar{\mathcal{S}} \setminus \{i_1, \dotso,i_N \}$ with $\bar{\bmh} \in C^2(\mathcal{S}) \cap C^{\infty}(\bar{\mathcal{S}} \setminus \{i_1, \dotso,i_N \})$. 
    \end{enumerate}}
    \label{Asymp-Euclidean-and-regular}
\end{definition}
\noindent  To use this definition in the case at hand, let us identify $(\mathcal{S},\bmh)$ with the $t=0$ hypersurface for the  metric \eqref{$q$-metric}, so that
\begin{equation*}
    \bmh =  - e^{2(\gamma_{q}-\psi_{q})} (\bmd{\varrho} \otimes \bmd{\varrho} + \bmd{z} \otimes \bmd{z}) - e^{-2 \psi_{q}} \varrho^2 \bmd{\phi} \otimes \bmd{\phi}.
\end{equation*}
Then, we introduce the coordinate transformation $(\varrho,z) \to (\bar{\varrho},\bar{z})$ defined by
\begin{equation*}
    \bar{\varrho} := \frac{\varrho}{\varrho^2 + z^2}, \qquad \bar{z} := \frac{z}{\varrho^2 + z^2}, \qquad \rho^{2}:= \varrho^{2} + z^{2} ,\qquad \bar{\rho}^2 := \rho^{-2},
\end{equation*}
and scale the metric $\bmh$ by the conformal factor $\bar{\Omega}= \rho^{-2} e^{\psi_{q} - \gamma_{q}} = \bar{\rho}^2 e^{\psi_{q} - \gamma_{q}}$ to get 
\begin{eqnarray*}
    && \bar{\bmh} = \bar{\Omega}^{2} \bmh, \\
    && \phantom{\bar{\bmh} } = - \bmd \bar{\varrho} \otimes \bmd \bar{\varrho} - \bmd \bar{z} \otimes \bmd \bar{z} -  e^{-2 \gamma_{q}} \bar{\varrho}^{2} \bmd \phi \otimes \bmd \phi. 
\end{eqnarray*}
The region at infinity is then identified with the points where $\bar{\Omega}=0$, which corresponds to $\rho \to \infty$ or $\bar{\rho}=0$. In fact, near $\bar{\rho}=0$, the conformal factor can be expanded as
\begin{equation}
    \bar{\Omega} = \bar{\rho}^{2} - m (1+q) \bar{\rho}^{3} + \mathcal{O}(\bar{\rho}^{4}),
    \label{Conformal-factor-expanded}
\end{equation}
where we have used the explicit form of $\psi_{q}$ and $\gamma_{q}$ given in \eqref{Potentials}. Given the above, we may identify the conformal extension $ \bar{\mathcal{S}} \cong \mathcal{S}_{t}  \cup \{ \infty \}$ and the point $i$ as the point where $\bar{\rho} =0$. Note that if we use coordinates $(\bar{\rho},\theta)$ instead of $(\bar{\varrho},\bar{z})$ in $\bar{\bmh}$, we get
\begin{equation}
    \bar{\bmh} = - \bmd \bar{\rho} \otimes \bmd \bar{\rho} - \bar{\rho}^{2} \bmd \theta \otimes \bmd \theta - \bar{\rho}^{2}  e^{-2 \gamma_{q}} \sin^{2} \theta \bmd \phi \otimes \bmd \phi,
    \label{three-dimensional-metric-spherical}
\end{equation}
where
\begin{equation*}
    \bar{\varrho} = \bar{\rho} \sin \theta,\qquad \bar{z} = \bar{\rho} \cos \theta.
\end{equation*}
The metric \eqref{three-dimensional-metric-spherical} is the standard $3$-dimensional flat metric in spherical coordinates, except for the $\bmd \phi \otimes \bmd \phi$ term. Given that
\begin{equation*}
    \bar{\rho}^{2} e^{-2 \gamma_{q}} = \bar{\rho}^{2} + \mathcal{O}(\bar{\rho}^{3}), \qquad \bar{\rho} \to 0,
\end{equation*}
the metric $\bmh$ can be written near $i$ in terms of Cartesian coordinates $\{ \bar{x}, \bar{y}, \bar{z} \}$, defined in the standard way, as
\begin{equation*}
    \bar{\bmh} = - \bmd \bar{x} \otimes \bmd \bar{x} - \bmd \bar{y} \otimes \bmd \bar{y} - \bmd \bar{z} \otimes \bmd \bar{z} + \mathcal{O}(\bar{\rho}^{3}),  \qquad \bar{\rho} \to 0.
\end{equation*}
Given $\bar{\Omega}=\bar{\rho}^2 e^{\psi_{q} - \gamma_{q}}$ and \eqref{Conformal-factor-expanded}, we can confirm that
\begin{equation*}
    \bar{\Omega}(i) =0, \qquad \bmd \bar{\Omega} (i) =0, \qquad \textbf{Hess } \bar{\Omega}(i) = - 2 \bar{\bmh} (i),
\end{equation*}
which verifies that $(\mathcal{S},\bmh)$ is asymptotically Euclidean and regular. 
{\small 
%\newpage
\bibliographystyle{unsrt} 
%\bibliography{refs} 
\newpage
%\iffalse

%\fi
\end{document}